\documentclass[useAMS,usenatbib]{mn2e}

\usepackage{amsmath}

\newcommand{\aap}{A\&A}

\newcommand{\apj}{ApJ}
\newcommand{\apjl}{\apj}

\newcommand{\mnras}{MNRAS}
\newcommand{\apjs}{ApJS}
\newcommand{\aapr}{A\&AR}
\newcommand{\pasj}{PASJ}
\newcommand{\araa}{ARA\&A}
\newcommand{\apss}{Ap\&SS}

\def\MSUN{\rm M_{\odot}}
\def\MDOT{\dot{M}}

\newbox\grsign \setbox\grsign=\hbox{$>$} \newdimen\grdimen
\grdimen=\ht\grsign \newbox\simlessbox \newbox\simgreatbox
\setbox\simgreatbox=\hbox{\raise.5ex\hbox{$>$}\llap
{\lower.5ex\hbox{$\sim$}}}\ht1=\grdimen\dp1=0pt
\setbox\simlessbox=\hbox{\raise.5ex\hbox{$<$}\llap
{\lower.5ex\hbox{$\sim$}}}\ht2=\grdimen\dp2=0pt
\def\simgreat{\mathrel{\copy\simgreatbox}}
\def\simless{\mathrel{\copy\simlessbox}}

\title[Magnetized Accretion Flows: Effects of Gas Pressure]{Magnetized Accretion Flows: Effects of Gas Pressure}
\author[M. Mo{\'s}cibrodzka and D. Proga]{M. Mo{\'s}cibrodzka$^{1}$\thanks{E-mail:
mmosc@illinois.edu} and D. Proga$^{2}$\\
$^{1}$ Department of Physics, University of Illinois, 1002 West Green Street, Urbana,
  IL 61801, USA\\
$^{2}$ Department of Physics, University of Nevada, 
4505 South Maryland Parkway, Las Vegas, NV
89154, USA}

\begin{document}

\date{Dates to be inserted}

\pagerange{\pageref{firstpage}--\pageref{lastpage}} \pubyear{2009}

\maketitle

\label{firstpage}

\begin{abstract}
We study how axisymmetric magnetohydrodynamical (MHD)  
accretion flows depend on $\gamma$ adiabatic index in the
polytropic equation of state.  This work is an extension of
Moscibrodzka \& Proga, where we investigated 
the $\gamma$ dependence of 2-D Bondi-like
accretion flows in the hydrodynamical (HD) limit.
Our main goal is to study if simulations for various
$\gamma$ can give us insights into to the problem of various modes of
accretion observed in several types of accretion systems such as
black hole binaries (BHB), active galactic nuclei (AGN),
and gamma-ray bursts (GRBs).  We find that for $\gamma \simgreat  4/3$, the fast
rotating flow forms a thick torus that is supported by rotation and gas
pressure. As shown before for $\gamma=5/3$, such a torus produces a strong,
persistent bipolar outflow that can significantly reduce
the polar funnel accretion of a slowly rotating flow. 
For low $\gamma$, close to 1, the torus is thin
and is supported by rotation. The thin torus produces an unsteady
outflow which is too weak to propagate throughout the polar funnel inflow.  
Compared to their HD counterparts, the MHD simulations show that the
magnetized torus can produce an outflow and does not exhibit regular
oscillations.  Generally, our simulations demonstrate how the torus thickness
affects the outflow production. They also support the notion that the
geometrical thickness of the torus correlates with the power of the torus
outflow.  Our results, applied to observations, suggest that the torus ability
to radiatively cool and become thin can correspond to a suppression of a jet
as observed in the BHB during a transition from a  hard/low 
to soft/high spectral state and a transition from a quiescent to 
hard/low state in AGN.
\end{abstract}


\section{Introduction}

Many astrophysical objects powered by accretion on to a compact objects like
e.g. active galactic nuclei -- AGN, black hole binaries -- BHB, gamma-ray
bursts -- GRBs, show signatures of outflows, either in a form of wind or jet,
or both.  Large scale, collimated jets are being detected from radio to
$\gamma$-ray wavelengths in AGN and microquasars (e.g. \citealt{konigl:2006},
\citealt{markoff:2006}).  Recently an increasing radio emission from the core
of some BHB was associated with small scale outflow/jet-like structures
\citep{fender:2004}.
The observations of BHB suggest that the power of 
the jet-like outflow may be related to the evolutionary stage of 
the object.

The BHB exhibit several X-ray spectral states, which can change in time
along a hysteresis loop on the luminosity-hardness diagrams. BHB (as well as
galactic nuclei) spend most of their time in the quiescent/off mode during
which their X-ray luminosity is low \citep{gallo:2003}. The discovery of 17
out of 20 known BHB was possible only during an outburst (X-ray novae)
accompanying the transition from the quiescent/off to a low/hard spectral
state after which the X-ray brightness increases three or more orders in
magnitude \citep{remillard:2006}. In the low/hard spectral state the
non-thermal power-law component dominates in the X-ray energy band.  After the
transition into the so-called high/soft state the spectrum is dominated by the
thermal multi-blackbody emission coming from a cold Keplerian disc
(\citealt{shakura:1973}).  The source returns to the quiescent state
through the soft-state, which closes the hysteresis cycle. It is also well
established that spectral states of BHB can change on relatively short
time-scales from days and months to years \citep{remillard:2006} and that the
X-ray lumonisty of the BHB does not necessarily correlate with its spectral
state \citep{done:2007}.

The spectral state changes
probably reflect the changes in the accretion rate/pattern and the
formation/suppression of an outflow.  For example, in the low/hard state, the
BHB radio luminosity is strongly correlated with the X-ray luminosity ($L_{\rm
R} \sim L_{\rm X}^{0.7}$, \citealt{gallo:2003}).  This correlation does not
hold after the transition to the high/soft state, when the radio emission
becomes suppressed (\citealt{corbel:2001}, \citealt{gallo:2003}).  The X-ray
variability and radio activity strongly support a model, where in the low/hard
state, a standard thin accretion disc is truncated and does not reach the
black hole horizon. During low/hard state, the inner disc that used to be
during the high/soft state is replaced by a hot thick flow, which probably
acts as a lunching site for a steady jet with the Lorentz factor, $\Gamma$,
typically less than 2. (e.g. \citealt{heinz:2003}; \citealt{meier:2005};
\citealt{remillard:2006}; \citealt{fender:2004} and references therein).

The transition from low/hard to high/soft spectral state occurs via
an intermediate state, the so-called very high state.
The very high state is characterized by
strong black body and substantial power-law emission.  
In the very high state, the radio emission exhibits
flaring. The flaring can be due to a fast non-stationary optically thin 
outflow with high Lorentz factor, $\Gamma >$ 2 \citep{fender:2004}.

A radio-X-ray correlation, similar to that seen in BHB, is also expected to
work in much larger scale systems like AGN.  Because of much longer
evolutionary time-scales in AGN, in comparison to BHB, the problem may be
studied statistically only.  The expectation of the correlation is supported
by the fact that the radio-loud quasars have higher X-ray fluxes than
radio-quiet quasars \citep{zdziarski:1995}.  In this scenario, the AGN
counterparts of the `quiescent' state are Low Luminosity AGN (LL AGN), because
they lack the 'big blue bump' in their spectra, are weak in X-rays and are
loud in radio in comparison to AGN (e.g. \citealt{ho:2003};
\citealt{kawakatu:2009}). Our own galactic center - Sgr A* - is considered as
an extreme example of LL AGN.  The models of the quiescent/off phase of AGN
and BHB, are usually based on the two assumptions: either most of the
accretion energy is transferred into the outflow kinetic energy which results
in a reduction of the mass-accretion rate, $\MDOT_{\rm a}$, and lower
luminosity, or the accretion is highly inefficient, or both
(\citealt{ichimaru:1977}, \citealt{narayan:1998},\citealt{yuan:2003}).

From the theoretical point of view, it is still unclear what induces the
accretion phase transition in BHB, AGN, and GRBs. Many authors considered 1-D, 
vertically integrated, models where the cold disc evaporates,
transitions into a hot phase and then re-condensates (e.g. \citealt{liu:1999},
\citealt{meyer:2000}, \citealt{rozanska:2000}, \citealt{liu:2002},
\citealt{meyer:2007}). In 1-D models, the mass-accretion rate is 
the key parameter responsible for the disc behaviour. These models 
include such physical processes as mass transfer between hot and cold phase, 
thermal conduction, and radiative transport.  However, the 1-D models do not 
include an outflow formation and do not yet attempt to reproduce observed 
correlation between X-rays and radio luminosity probably caused by outflow, 
during low/hard state.
As underlined in \citet{abramowicz:2000}, at least a 2-D treatment of 
the problem
is needed to correctly model the accretion state transition process.

In this paper, we present results of 2-D axisymmetric magnetohydrodynamical
(MHD) simulations of a slowly rotating accretion flow on to a black hole
(BH).  We investigate the time evolution and properties of inflows/outflows
for various values of an adiabatic index, $\gamma$.  Our work is an extension
of work of \citet{moscibrodzka:2008} (hereafter MP08) who considered models
for various $\gamma$ but in the hydrodynamical (HD) limit (see also
\citealt{proga:2003a}, hereafter PB03a, for HD simulations with $\gamma=5/3$,
work of \citealt{proga:2003b}, hereafter PB03b, who considered MHD models only
for $\gamma=5/3$, and finally work of \citet{proga:2003} who considered a MHD
collapsar model for GRBs where the inner accretion flow is radiation pressure
dominated and $\gamma$ is effectively $\approx 4/3$).

In MP08, we showed that in the HD limit, many properties of the accretion flow
strongly depend on $\gamma$. Most relevant to this work is that the thickness
of the torus decreases with decreasing $\gamma$ whereas $\MDOT_{\rm a}$ in
units of the Bondi rate, $\MDOT_{\rm B}$ increases with decreasing
$\gamma$. Here, we want to check how $\gamma$ affects magnetized flows. For
example, is the efficiency of producing outflows sensitive to the thickness of
the torus? This question is related to the observations of BHBs and AGN.  One
could argue that magnetically driven outflows should be the strongest from
geometrically thick discs/torii with significant pressure support and the
weakest, if any, from geometrical thin discs with little pressure support
(e.g. \citealt{livio:1999}; \citealt{meier:2001}).

To our best knowledge, there are no direct simulations that would confirm the
above expectation mainly because  previous simulations considered thick,
pressure/rotation supported torii. What is clear is that
most of these previous global MHD
simulations of accretion flows show outflows (e.g. \citealt{uchida:1985};
\citealt{stone:1994}; \citealt{stone:2001}; \citealt{hawley:2002}; PB03b;
\citealt{proga:2003};\citealt{devilliers:2005}; \citealt{mizuno:2004};
\citealt{kato:2004}; \citealt{mckinney:2004}; \citealt{proga:2005},
\citealt{fragile:2009}). However, it is unclear under what conditions 
(e.g., for what torus thickness) the torus stops
to produce an outflow.

Our very simplistic approach to control the torus thickness
is dictated by a complexity of the relevant physics.
A full treatment of an accretion disc, especially
geometrically thin one, requires modeling energy dissipation, radiative
heating and cooling, and radiative transfer. In our approach, for $\gamma$=
5/3 and 4/3 the torus will be hot and supported against gravity by gas
pressure and rotation. As such it will be geometrically thick. On the other
hand, for $\gamma \sim 1$ the torus will be supported only by rotation
and therefore it will be Keplerian and geometrically thin (or at least
considerably thinner than for $\gamma=5/3$). Therefore, our study of effects
of $\gamma$ will allow us to gain some insights into the problem of the
outflow formation for various physical conditions and different phases of
activity like e.g. soft and hard spectral states in BHB and AGN activity modes.

Here we will not be able to immediately compare our results with observations
in any great detail (as it was possible in \citealt{moscibrodzka:2007} where
we computed and compared radiative properties predicted by
PB03b's MHD simulations
with observations of Sgr A*). However, our approach enable us to produce
results that can be easily and
consistently compared with a great body of data from complementary numerical
simulations that were preformed within one framework and with the same
numerical tools (i.e.,PB03a; PB03b; \citealt{proga:2003};
\citealt{proga:2005}, MP08).

The article is organized as follows. In \S~\ref{sec:2} we describe our method 
and initial conditions used to set our simulation.
In \S~\ref{sec:3} we present the results of the numerical
MHD simulations. The results are discussed in the contest of
previous work in \S~\ref{sec:4}.

\section{Method}
\label{sec:2}
\subsection{Equations}

To calculate the structure and evolution of an accreting flow, we
solve the equations of MHD
\begin{equation}
   \frac{D\rho}{Dt} + \rho \nabla \cdot \boldsymbol{v} = 0,
\label{eq:con}
\end{equation}
\begin{equation}
   \rho \frac{D \boldsymbol{v}}{Dt} = - \nabla P + \rho \nabla \Phi +
   \frac{1}{4 \pi}(\boldsymbol{\nabla} {\bf \times} \boldsymbol{B}) \times \boldsymbol{B},
\label{eq:mom}
\end{equation}
\begin{equation}
   \rho \frac{D}{Dt}\left(\frac{e}{ \rho}\right) = -P \nabla \cdot
   \boldsymbol{v},
\label{eq:en}
\end{equation}
\begin{equation}
  \frac{\partial \boldsymbol{B}}{\partial t}=\boldsymbol{\nabla} {\bf \times}
( \boldsymbol{v}  {\bf \times} \boldsymbol{B}),
\label{eq:mag}
\end{equation}
where $\rho$, {\it P}, $\boldsymbol{v}$, $e$, $\Phi$ and $\boldsymbol{B}$ is
the mass density, gas pressure, fluid velocity vector, internal energy density,
gravitational potential, and magnetic field vector, respectively. 
  Lagrangian derivative is defined as: $\frac{D}{Dt}=\frac{\partial}{\partial t} +
  \boldsymbol{v} \cdot \boldsymbol{\nabla}$.
We relate $P$ and $e$ through the equation of state
$P~=~(\gamma-1)e$, where $\gamma$
is an adiabatic index.

We perform simulations using the pseudo-Newtonian potential $\Phi$ 
introduced by \citet{paczynski:1980} (here after PW)
\begin{equation}
\Phi=-\frac{G M}{r-R_S},
\end{equation}
where $R_{\rm S}$ is the Schwarzschild radius.
This potential approximates general relativistic effects in the inner
regions, for a non-rotating BH.  In particular, the
PW potential reproduces the last stable circular
orbit at $r=3 R_{\rm S}$ as well as the marginally bound orbit at $r=2 R_{\rm S}$.

\subsection{Initial conditions and boundary conditions}
\label{sec:init_mhd}

Generally, the setup and numerical methods used in our
simulations  are as in PB03b.
Here, we only briefly summarize the initial and boundary conditions 
and spell out
the differences between our and PB03b simulations.

Our calculations are performed in spherical
polar coordinates $(r,\theta,\phi)$. We assume axial symmetry about
the rotational axis of the accretion flow ($\theta=0^\circ$ and
$180^{\circ}$).
For the initial conditions of the fluid variables we follow PB03b and
adopt a Bondi accretion flow.  In particular, we adopt $v_r$ and
$\rho$ computed using the Bernoulli function and mass accretion rate
for spherically symmetric Bondi accretion with the PW potential.  We
set $\rho_{\rm \infty}=2.2\times 10^{-23}$ $\rm{g/cm^3}$ and mass of the
central black hole $M_{\rm BH}= 3.7\times 10^6\MSUN$
which make our results directly applicable to 
the galactic center (e.g. it will be useful in our future 
calculations of synthetic spectra
as in \citealt{moscibrodzka:2007}). However,
our results could be rescaled and applied to other systems.
We specify the sound speed at infinity, $cs_{\rm \infty}$
through a normalized Schwarzschild radius 
$R'_{\rm S}\equiv R_{\rm S}/R_{\rm B}$ ($R'_{\rm S}=2cs^2_{\rm \infty}/c^2)$,
where $R_{\rm B}$ is the Bondi radius.
The normalized BH radius
characterizes the gas temperature.  We consider models with $R'_{\rm S}=10^{-3}$.

We specify the initial
conditions by adopting a non-zero specific angular momentum {\it l} at
the outer boundary. We consider a case in which
the angular momentum at the outer radius
$r_{\rm o}$ depends on the polar angle via
\begin{equation}
l(r_o,\theta)=l_0 (1-|\cos\theta|).
\end{equation}
This formula was used by PB03a and MP08.
We note that PB03b use a step function so
that this is one of the differences between the setup of our and PB03b's
simulations. However, as discussed by PB03b and shown below
the details of the {\it l} distribution are not very important.
We express the angular momentum on the equator as
\begin{equation}
l_0=\sqrt{R'_C} R_B cs_\infty,
\end{equation}
where $R'_{\rm C}$ is the `circularization radius' on the equator in units
of $R_{\rm B}$ for the Newtonian potential (i.e., $GM/r^2= v^2_\phi/r$ at
$r= R'_{\rm C} R_{\rm B}$).

To generate the magnetic field, 
we use vector potential, $\boldsymbol{A}$, i.e., $\boldsymbol{B=
\nabla \times A}$.  The initial magnetic field configuration is
a pure vertical magnetic field. The zero divergence 
vertical field is defined by the potential 
$\boldsymbol{A}$=($A_r=0$, $A_{\theta}=0$, $A_{\phi}=A r \sin \theta$).
We scale the magnitude magnetic field by plasma parameter $\beta_{\rm in}=8 \pi
P/B^2$ at the inner boundary, $r_{\rm i}$.
\begin{equation}
A=\sqrt{(P_{r_i} / \beta_{in})},
\end{equation}
where $P_{\rm r_i}$ is the initial pressure at the inner radius.

Our standard computational domain is defined to occupy the radial range
$r_{\rm i}~=~1.5~R_{\rm S} \leq r \leq \ r_{\rm o}~=~ 1.2~R_{\rm B}$ and the angular range $0^\circ
\leq \theta \leq 180^\circ$. The
$r-\theta$ domain is discretized into zones with 140 zones in the $r$
direction and 100 zones in the $\theta$ direction.  We fix zone size ratios,
$dr_{k+1}/dr_{k}=1.05$, and $d\theta_{l}/d\theta_{l+1} =1.0$ for $0^\circ \le
\theta \le 180^\circ$.  With such defined grid we are able to resolve the
  fastest growing mode of the magnetorotaitonal instability (MRI) only in a
  case of thick torus ($\gamma$=5/3). See \S~\ref{sec:structure}.

The boundary conditions are specified as follows. At the poles, (i.e.,
$\theta=0^\circ$ and $180^\circ$), we apply an axis-of-symmetry
boundary condition. At both the inner and outer radial boundaries, we
apply an outflow boundary condition for all dynamical variables.  As
in PB03b, to represent steady conditions at the outer radial boundary,
during the evolution of each model we continue to apply the
constraints that in the last zone in the radial direction,
$v_\theta=0$, $v_\phi=l_0 f(\theta)/ r_{\rm o} \sin{\theta}$ 
(where $f(\theta)=1-|\cos\theta|$), and the density
is set to $\rho_\infty$ at all times. Note that we allow $v_r$ to
float. For the magnetic field, at the outer boundary, we apply an outflow 
condition.
At the inner boundary, we follow \citet{stone:2001} 
and use negative stress condition
(i.e. we enforce $B_{r} B_{\phi} \leq 0$ at $r_{\rm i}$).
To reduce the problem of high Alfv{\'e}nic velocities in regions of low density,
we set a lower limit to the density on the grid as $\rho_{\rm min}=\rho(r_{\rm
  i}/r)^{1/2}$
and enforce it at all times in the simulations. 

To solve Eqs.~(\ref{eq:con})-(\ref{eq:mag}) we use the {\tt zeus-2D} code
described by \citet{stone:1992a,stone:1992b}, modified to implement the
PW potential.

\section{Results of the simulations} 
\label{sec:3}
The parameters used for the simulations are summarized in
Table~\ref{tab:mhd_sum}. The Table columns (1)-(8) show respectively 
the number of a given run; 
the BH radius and the Bondi radius ratio ($R'_{\rm S}$);
the circularization radius compared to the Bondi radius ($R'_{\rm C}$); 
the adiabatic index ($\gamma$); 
the initial magnetic plasma parameter ($\beta_{\rm in}$) at $r_{\rm o}$; 
the final time ($t_{\rm f,in}$) at which we stopped each simulation 
in units of the Keplerian orbital time at $r_i$ ($t_{\rm dyn} = 595 s$ for 
$M_{\rm BH}= 3.7\times 10^6\MSUN$); 
the same final time $t_{\rm f,out}$ but in units
of the Keplerian orbital time at $r_{\rm o}$, 
and finally the time-averaged mass 
accretion rate on to the BH in units of the Bondi accretion rate 
($\MDOT_{\rm a}/\MDOT_{\rm B}$) over 1000 $t_{\rm dyn}$ at the end of each run.
The shorter duration of our simulations (compared to the duration of PB03b's 
simulations) is caused by the high density gradients
which appear in our runs for low $\gamma$, 
and which lead to significantly smaller time steps.

\begin{table*} \footnotesize
\begin{center}
\caption{ Summary of parameter survey of the MHD simulations.}
\begin{tabular}{l c c c c c c c c l } \\ 
\hline & & & & & & 
\\ run & $R'_{\rm S}$ & $R'_{\rm C}$ & $\gamma$ & $\beta_{\rm in}$ & $t_{\rm
  f,in}$ & $t_{\rm f,out}$ & $\MDOT_{\rm a}/\MDOT_{\rm B}$ \\

\hline 
     1 & $10^{-3}$ & $5\times10^{-2}$& 5/3 &$ 10^9$ & $1.22 \times 10^4$ & 0.55 & 0.03 \\ 
     2 & $10^{-3}$ & $5\times10^{-2}$& 4/3 &$  10^9$ & $1.2 \times 10^4$ & 0.54 & 0.01 \\ 
     3 & $10^{-3}$ & $5\times10^{-2}$& 1.2 &$ 10^9$ & $8.76 \times 10^3$ & 0.4  & 0.1 \\ 
     4 & $10^{-3}$ & $5\times10^{-2}$& 1.01 &$ 10^9$& $1.36 \times 10^4$ & 0.62 & 0.35 \\

\hline
\end{tabular}
\label{tab:mhd_sum}
\end{center} \normalsize
\end{table*}

\subsection{Velocity field and sonic surfaces evolution}

In our simulations, the evolution of the poloidal velocity field initially
proceeds in a similar manner as in the corresponding HD simulations (see
e.g. panels 3 and 4 in fig.2 in PM08 for each $\gamma$).  Namely, after a
short episode of the radial inflow, the matter with relatively high-{\it l}
forms a torus in the equatorial plane around the BH
(high-{\it l} is defined here as $l > l_{cr}$ and low-{\it l} as $l < l_{cr}$, 
where $l_{cr} = 2 R_{\rm S}c$ is a critical specific angular momentum).
For $\gamma$=5/3, 4/3,
and 1.2, initially when the torus forms, a shock wave is produced.  
The shock wave forms because some of the gas forming a torus turns around 
and moves outward along
the equator due to the centrifugal force and gas pressure or tries to move
closer to the poles and accrete.  For $\gamma=$1.01, the gas is the most
gravitationally bound compared to other $\gamma$'s.  As such it accumulates
near BH and does not flow out along the equator contrary to the higher
$\gamma$ cases.  

In its early phase of the evolution, 
the torus is not able to accrete on to a BH, independently of
$\gamma$, because the MHD turbulence has not yet fully developed to 
transoport the
angular momentum. The matter in the polar regions does not feel any centrifugal
barier, thus it flows radially toward the BH. In our MHD simulations and
the HD simulations studied in PM08, the sonic surfaces topology evolves in a
similar manner. It changes from a circular shape into two lobes where the
lobes are elongated along the poles. As we reported in PM08, the evolution of
the sonic surface is the slowest for $\gamma$=1.01, due to a lack on an
equatorial outflow.

After a short time the magnetic field effects start to dominate the accretion
flow evolution. The torus starts to accrete toward the BH.
The biggest and expected difference between HD and MHD
simulations is the development of bipolar outflows in MHD simulations
regardless of $\gamma$.  PB03b's simulations well illustrate the general
evolution of the magnetic fields. The polar outflows form at
small radii ($r < 20 R_{\rm S}$) close to the poles and expand to the outer
boundary of the computational domain along the symmetry axis. We find that the time at
which the outflows form depends on $\gamma$: the lower $\gamma$, the later the
outflow forms. Although initially HD and MHD flows evolve similarly, at the
end of the simulations, the geometry and character of the accretion flows in
the HD and MHD simulations, are significantly different. At the end of the
simulation, the MHD flows are composed of a bipolar outflow propagating in the
polar regions, and the inflow near the equator.  The equatorial region is
turbulent and strongly asymmetric for all $\gamma$'s.

\subsection{Mass accretion rate evolution} 

\begin{figure*}
\begin{picture}(0,600)

\put(-100,300){\includegraphics{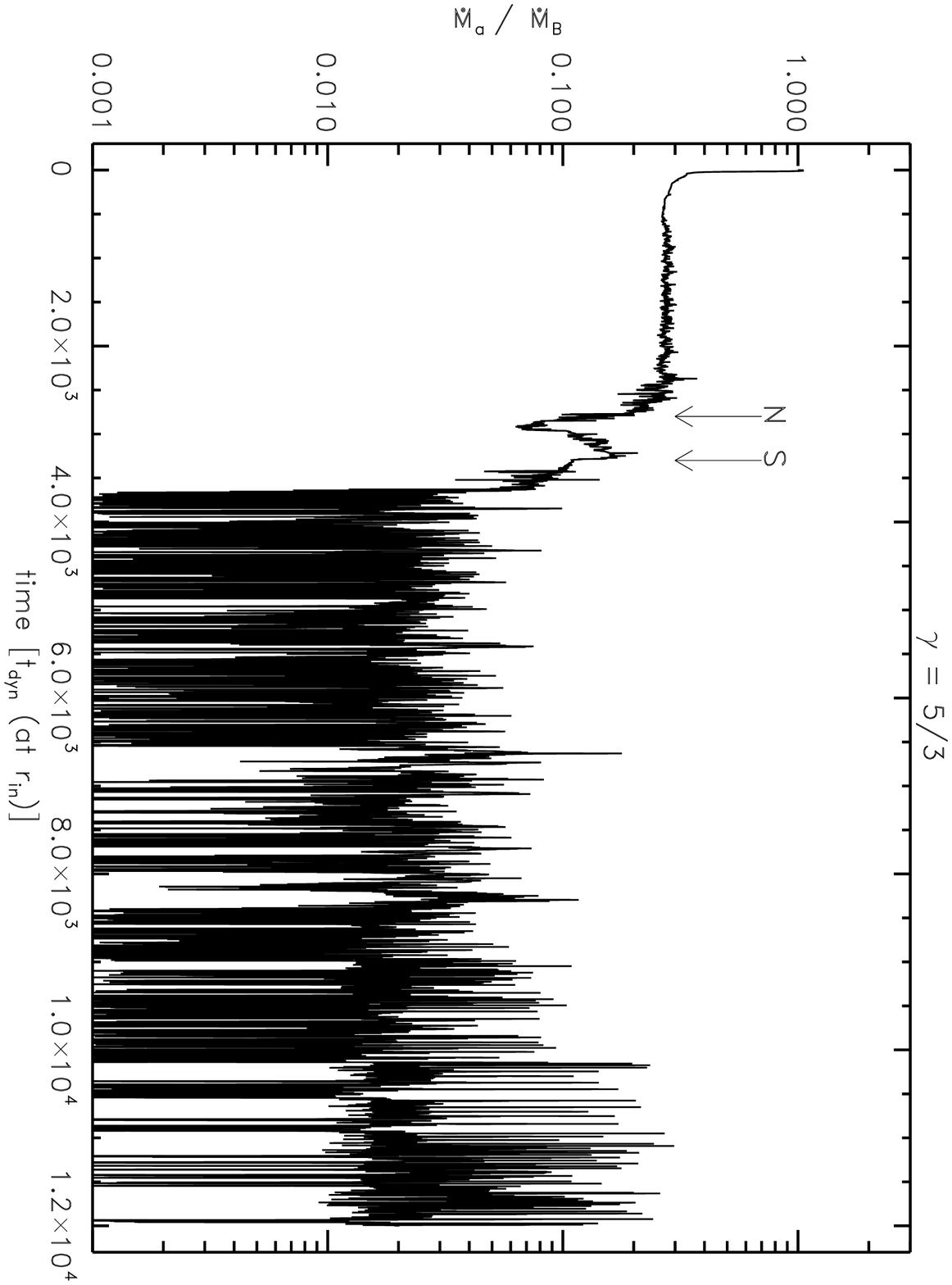}}

\put(140,300){\includegraphics{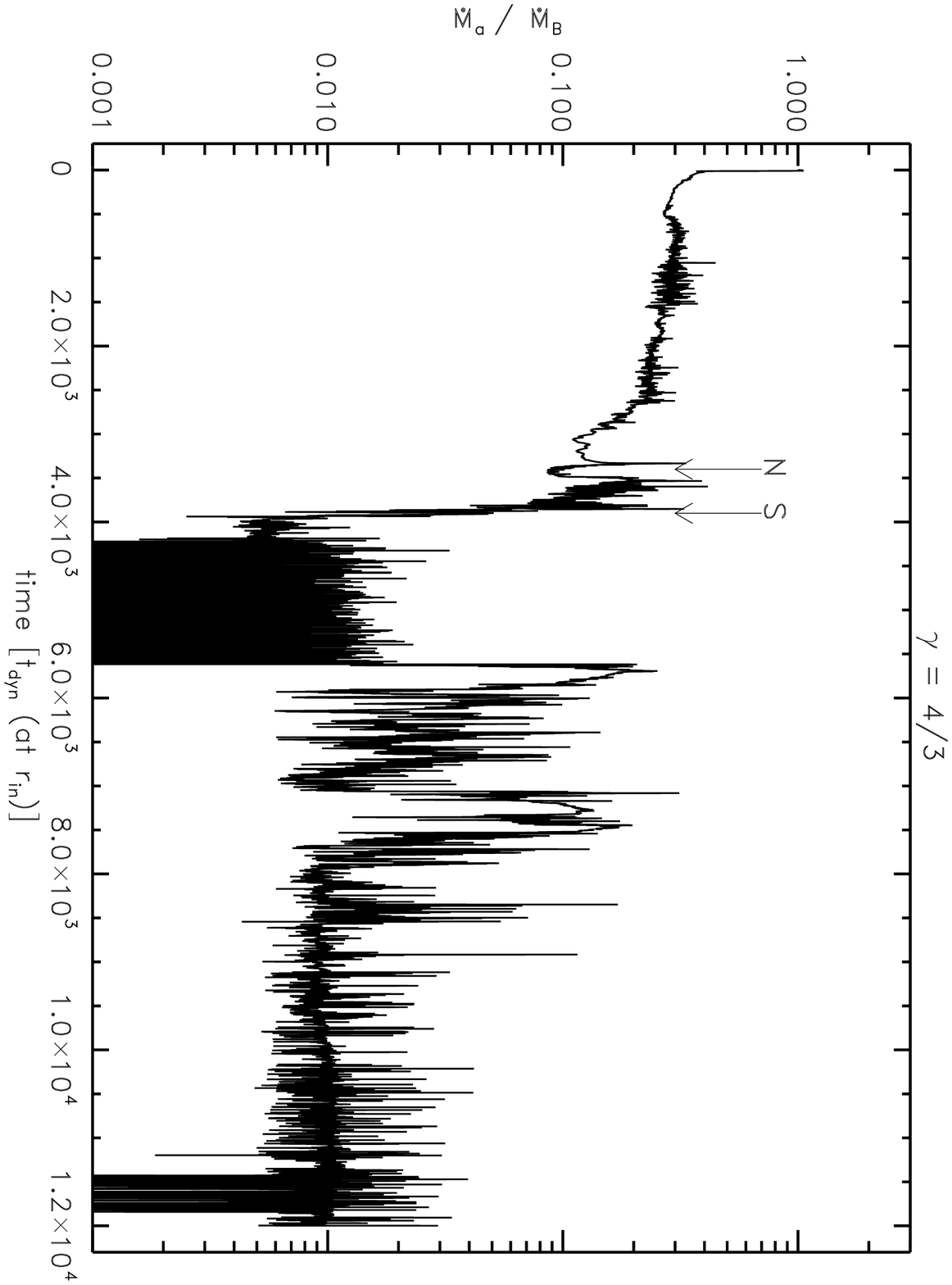}}

\put(-100,100){\includegraphics{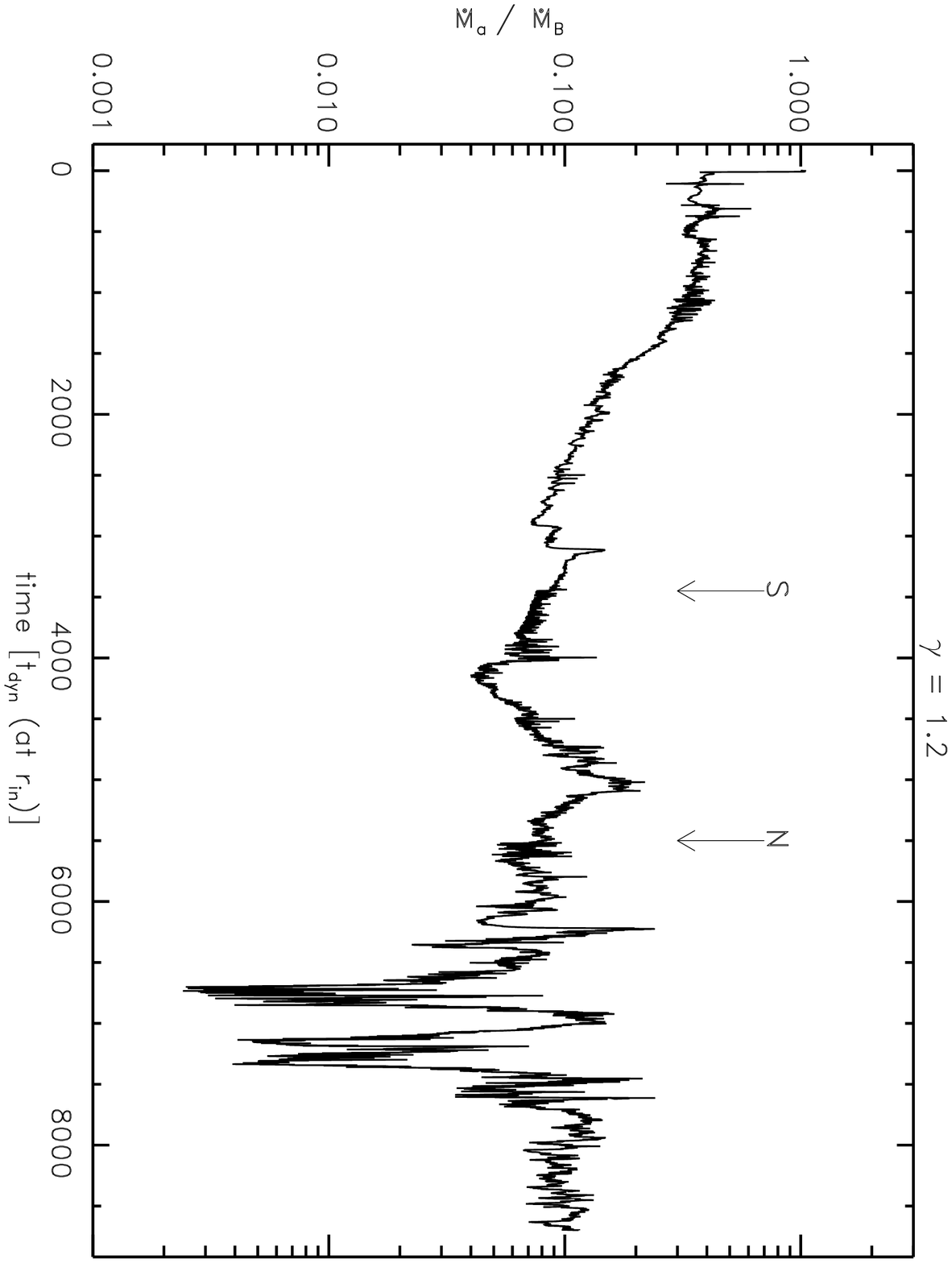}}

\put(140,100){\includegraphics{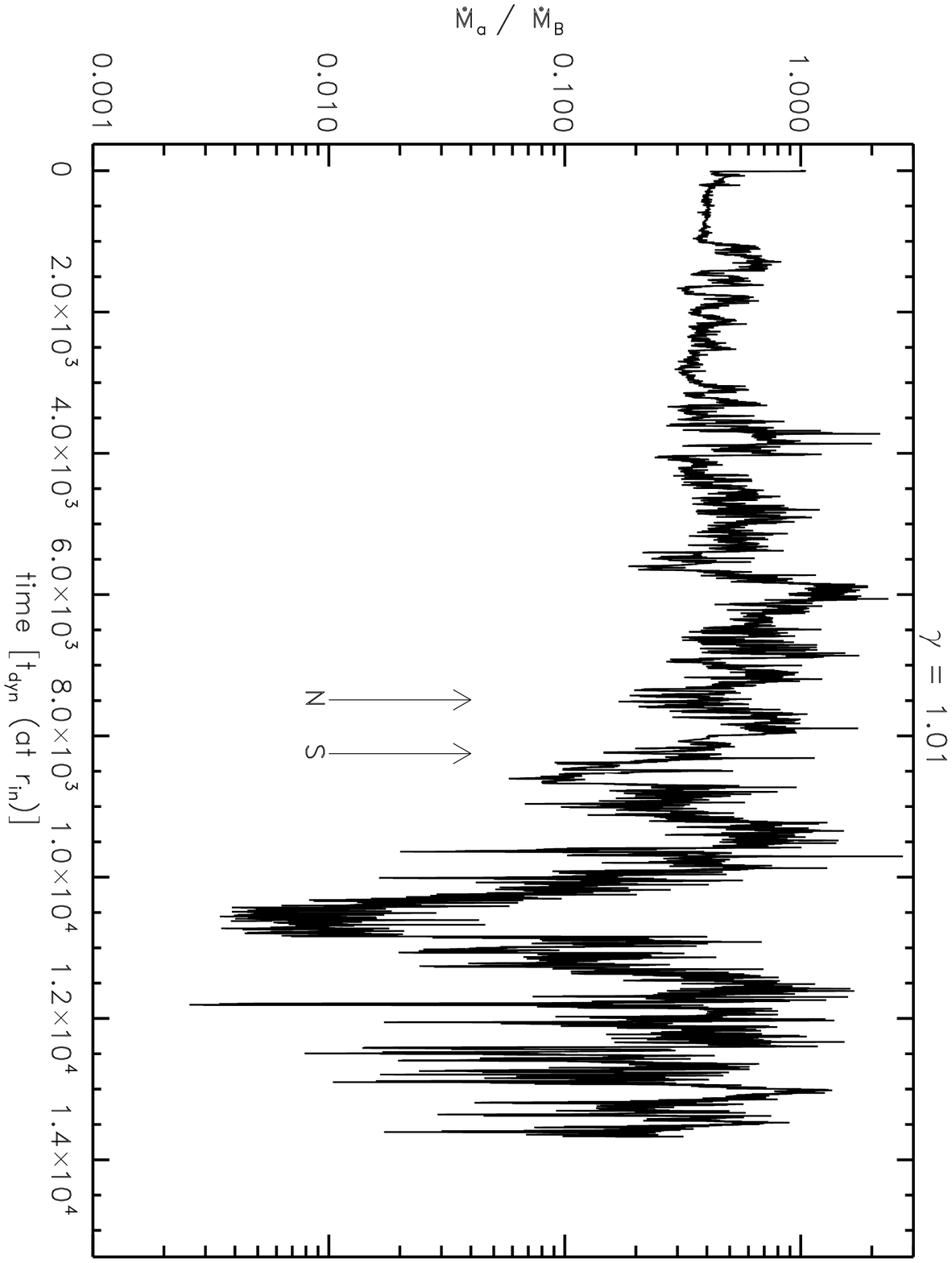}}

\end{picture}
\caption{Each panel shows the evolution of the mass accretion rate for the
models in Table~\ref{tab:mhd_sum} with different $\gamma$ parameter. $\gamma$=5/3
(upper-left), 4/3 (upper-right), 1.2 (lower-left), and 1.01 (lower-left). The
arrows N and S indicate the moments when the polar outflow forms in the
Northern and Southern direction, respectively. The time is given in units of
dynamical time at the inner boundary of the flow $t_{\rm dyn}$=595 s.}
\label{fig:accretion_evolution}
\end{figure*}

Fig.~\ref{fig:accretion_evolution} shows the
$\MDOT_{\rm a}/\MDOT_{\rm B}$ as a function of time, for each studied $\gamma$ as listed in
Table~\ref{tab:mhd_sum}. The time is given in units of the orbital time at the
inner boundary of the computational domain.  A pair arrows in each panel 
mark the moments at which the polar outflow forms, 
in the Northern (N) and Southern (S) directions. 
The moment of the formation of the large scale polar outflow is 
defined as a moment when the matter stops accreting 
through the polar regions and the radial velocity sign changes sign from
negative to positive. 
We determine this by inspecting the velocity maps
and checking that the outflow velocity becomes larger than
escape velocity at the outer boundary.

The upper-left panel in Fig.~\ref{fig:accretion_evolution} shows $\dot{M}_{\rm
a}/\dot{M}_{\rm B}$ for run~1.  Initially, $\MDOT_{\rm a}/\MDOT_{\rm B}$
decreases from 1 to 0.3 and then remains constant until $t \approx $ 2500.
The value for $\dot{M}_{\rm a} /\dot{M}_{\rm B}$ of 0.3 is consistent with
that obtained in the HD counterpart (see MP08 and PB03b), and it corresponds
to the matter accreting in the polar funels in the initial HD phase
of the simulation. However, later around t= 3700, $\dot{M}_{\rm a}/
\dot{M}_{\rm B}$ suddenly decreases to a minimum value (below 0.001), and
oscillates between the minimum value and 0.03. We note that the moment of the
S outflow formation does not correspond exactly to the moment of the
$\MDOT_{\rm a}/\MDOT_{\rm B}$ reduction to its minimum value. In general (the
details will be discussed in the next sections) the mass accretion rate
reduction is caused by the mentioned bipolar outflow, which pushes the
radially infalling material away from the polar funels.  Such a vacuuming of the
polar regions makes the torus the only source of gas available for
accretion. During later phases of the evolution (i.e., $t > 4000$),
$\MDOT_{\rm a}/\MDOT_{\rm B}$ shows narrow spikes and dips similar to those
found by PB03b, corresponding to the episodes of enhanced and suppressed
torus accretion, respectively.
  
The upper-right panel in Fig.~\ref{fig:accretion_evolution} shows the
$\MDOT_{\rm a}/\MDOT_{\rm B}$ evolution for $\gamma$=4/3 (run 2).  At the
beginning but after the torus formation, $\dot{M}_{\rm a} / \dot{M}_{\rm
B}\approx$ 0.2, which is consistent with mass accretion rate found previously
in the HD simulations. At this stage, the gas is being accreted from the polar
regions.  Later when an outflow forms at $t\approx 4000$, $\MDOT_{\rm
a}/\MDOT_{\rm B}$ suddenly decreases.  For $t >4000$, the $\MDOT_{\rm
a}/\MDOT_{\rm B}$ evolution is characterized by narrow spikes, dips (similarly
to run 1), and additionally broader features with higher amplitudes. The mass
accretion rate in the spikes can reach a level of about 0.02 and corresponds to
the torus accretion.  During the dips, the $\dot{M}_{\rm a}/ \dot{M}_{\rm B}$
has a minimum value below 0.001, when the torus stops accreting.  The rates
seen in the two broad features (around t=5500 and t=7000) reach the level of
$\sim$ 0.1. At even later times ($t > 9000$), $\MDOT_{\rm a}/\MDOT_{\rm B}$ is
almost constant around 0.01. The broad features have been found before by
PB03b, and correspond to the non-rotating gas accretion that reenter the BH
vicinity in the equatorial plane.  Near the end of the simulation, the minima 
(dips) reappear.

For run 3 (the lower-left panel in Fig.~\ref{fig:accretion_evolution}), the
evolution of the mass accretion rate differs significantly from runs 1 and 2.
In particular, there are no sudden drops in $\dot{M}_{\rm a}/\dot{M}_{\rm
B}$. Instead, $\MDOT_{\rm a}/\MDOT_{\rm B}$ evolves quite smoothly, despite
the formation of bipolar outflows.  The $\dot{M}_{\rm a} / \dot{M}_{\rm B}$
curve is variable, and it shows two significant minima around t=7000 and at
the end of the simulation. The mean level of $\dot{M}_{\rm a} / \dot{M}_{\rm
B}$ is around 0.1, which is rather high compared to those of runs 1 and 2.

The lower-right panel in Fig.~\ref{fig:accretion_evolution}, presents
$\dot{M}_{\rm a} / \dot{M}_{\rm B}$ for the model with $\gamma$=1.01 (run
4). Here, $\MDOT_{\rm a}/\MDOT_{\rm B}$ shows a strong time variability
although the time averaged rate remains high compared to other models. There
is one significant minimum in $\dot{M}_{\rm a} / \dot{M}_{\rm B}$ at the final
stages of the simulation.  The panel shows many periods of increased
accretion, during which $\MDOT_{\rm a}/\MDOT_{\rm B}$ is higher than 1!

Interestingly, for $\gamma$=1.2 and 1.01, in spite of the presence of the
bipolar outflows, $\MDOT_{\rm a}/\MDOT_{\rm B}$ does not decrease as sharply
as for
 $\gamma$=5/3 and 4/3. It also does not show the characteristic
features such
 as spikes and dips which were also found by PB03b.  The
different character of the mass accretion rate evolution in runs: 1,~2 and
3,~4, reflects two distinct outflow formation mechanisms.

\subsection{Inflow/outflow rates}

To investigate the mechanisms responsible for the polar outflow formation, 
we examine inflow/outflow/total fluxes of various physical
quantities as functions of the distance from BH.  The fluxes are
averaged over the whole $\theta$ range and over 1000 $t_{\rm dyn}$
before the end of each run.

We compute the fluxes of: mass, internal energy, kinetic energy,
magnetic energy and total energy. 
The mass flux  at a given radius, $r$ 
is given by:
\begin{equation}
\dot{M}= r^2 \oint_{4 \pi} \rho v_r d\Omega
\label{eq:mass}
\end{equation}
where $d\Omega = \sin(\theta) d\theta d\phi$, $\rho$ is the density,
and $v_{r}$ is the radial velocity.  To calculate the mass inflow rate
$\dot{M}_{\rm in}$ at a given radius, we include only contributions with
$v_r < 0$, whereas to calculate the mass outflow rate $\dot{M}_{\rm out}$,
we include only contributions with $v_r > 0$. To
compute the net (total) mass flow flux $\dot{M}_{tot}$, we include all
contributions regardless of the $v_r$ sign 
( $\dot{M}_{tot} = \dot{M}_{\rm in} +  \dot{M}_{\rm out}$). 
To compute  the internal, kinetic, magnetic and total energy fluxes,
we use
the same approach as for the mass fluxes with small modification. Namely,
$\rho$ in Eq.~\ref{eq:mass} is substituted by {\it e}, $E_{\rm kin}=\rho
(v_r^2+v_{\theta}^2 + v_{\phi}^2)/2$, $E_{\rm Poynting}=(\boldsymbol{v} \times
\boldsymbol{B})/v_r $ and $E_{\rm tot} = E_{\rm kin} + e + E_{\rm Poynting}$ for the
internal, kinetic, magnetic and total energy, respectively.  We
normalize all energy fluxes with $\dot{M}_{\rm B}c^2$.

\begin{figure*}
\begin{picture}(0,600)

\put(120,0){\includegraphics{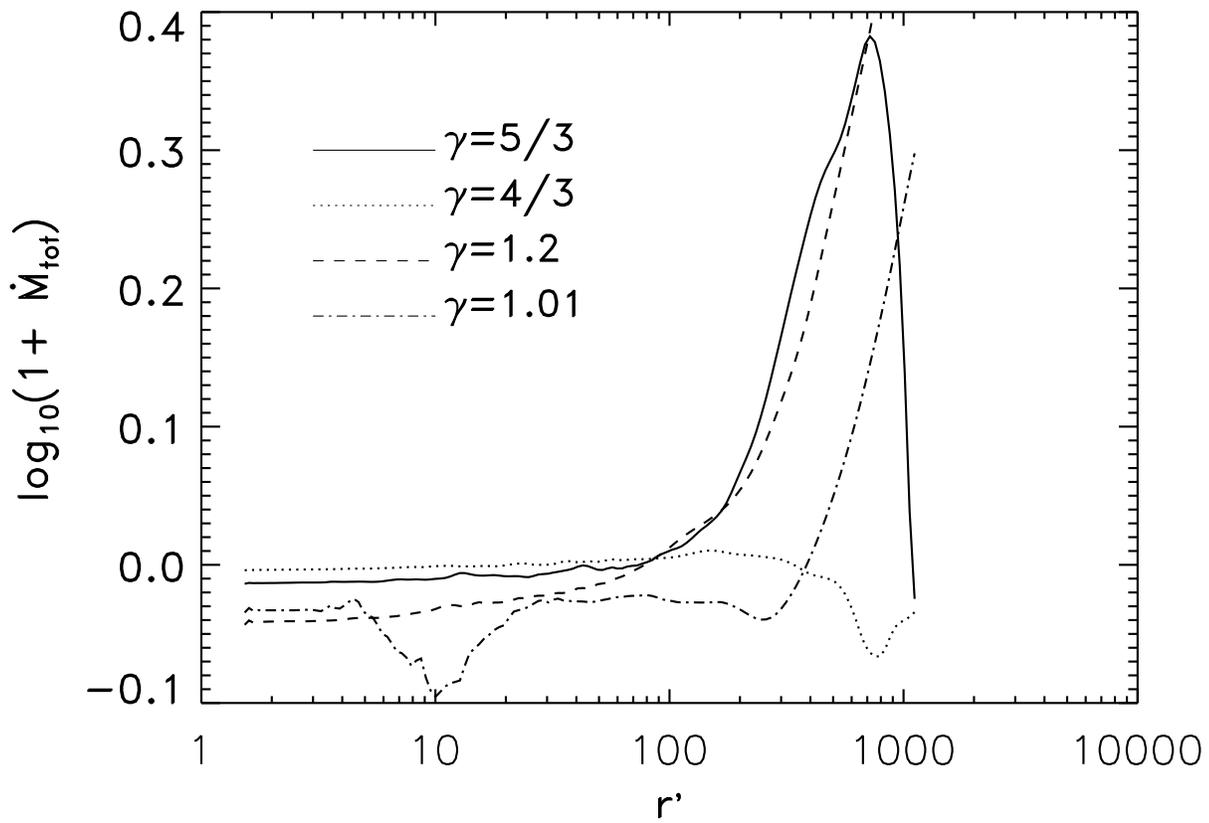}}

\end{picture}
\caption{ The total mass flux, $\dot{M}_{tot}$, as a function of radius
for $\gamma$=5/3 (solid), 4/3
(dotted), 1.2 (dashed), 1.01 (dot-dashed line). 
}
\label{fig:mdot_in_out_tot}
\end{figure*}

\begin{figure*}
\begin{picture}(0,600)

\put(120,0){\includegraphics{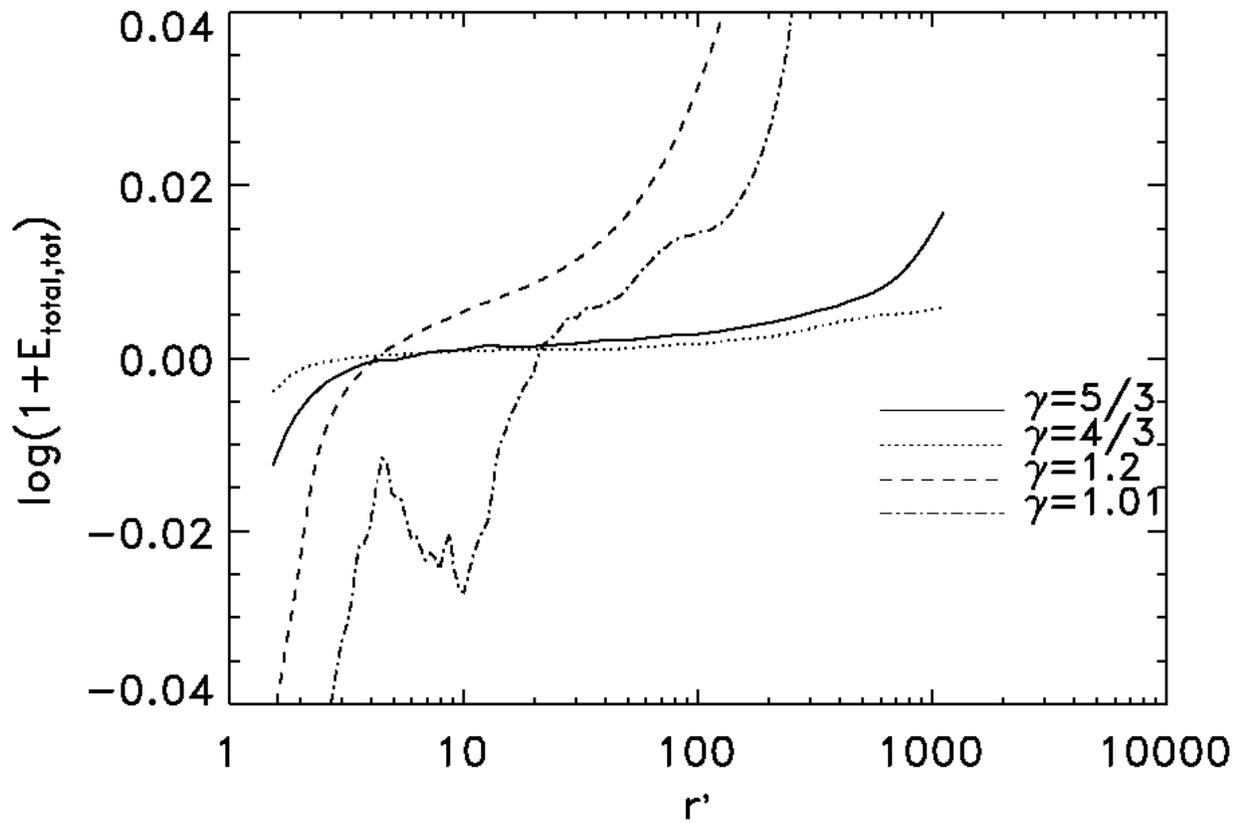}}

\end{picture}
\caption{The total flux of total energy
as a  function of radius for $\gamma$=5/3 (solid), 4/3 (dotted), 1.2
  (dashed), 1.01 (dot-dashed line).}
\label{fig:Etot_in_out_tot}
\end{figure*}

In Fig.~\ref{fig:mdot_in_out_tot} and Fig.~\ref{fig:Etot_in_out_tot}, 
we only show the total mass flux and total energy fluxes, respectively.
We present it as a $log(1+\dot{M}_{tot}$) function so that the positive
  values will indicate the domination of an outflow and the negative - an inflow.
The solid, dotted, dashed and dot-dashed lines
correspond to runs 1, 2, 3, and 4, respectively (as marked in the panels).
Below, we will also the main conclusion which will be based
our analysis of all studied fluxes not just those presented in the figures.

The total mass flux (Fig.~\ref{fig:mdot_in_out_tot}) 
as a function of radius is not constant. This indicates 
that overall the flows did not reach a steady state. However, 
in run 1 and 2, the total mass flux is
constant in the inner parts of the computational domain only up to 
few $R_{\rm S}$.  For models 3 and 4, the flux  is nearly constant over a wider
radial range, i.e., up to $r \sim 100 R_{\rm S}$.

Our analysis of the various fluxes showed that
in run 1, the total internal energy flux changes sign from negative (domination
of the inflow flux) to positive (domination of the outflow flux) very close
to the BH ($r \sim 10 R_{\rm S}$). 
The similar behaviour characterizes run 2
and 3.  Run 4 differs from other cases:  its total internal energy flux
is negative for all radii. 

There are a few general trends in the behaviour of all fluxes in all runs.
The strongest contribution to the total energy inflow flux (upper panel in 
Fig.~\ref{fig:Etot_in_out_tot}) in the outer region of
the computational domain (which we define here as the region for 
$r \ge 10 R_{\rm S}$), is from the internal energy, 
regardless of the value of $\gamma$, whereas the total
energy outflow flux (middle panel in Fig.~\ref{fig:Etot_in_out_tot}) 
is dominated by kinetic energy (with one exception for
$\gamma=1.01$ for which the internal energy outflow flux is even greater than
kinetic one but only up to $r \sim 100 R_{\rm S}$).

In the inner part of the flows (for $r \le 10 R_{\rm S}$), where the
most of the outflow is produced, the total energy inflow flux is 
dominated by the kinetic energy for all runs.

 The total energy outflow flux in the close vicinity to the BH (i.e. for $r \le
10 R_{\rm S}$) is dominated by magnetic energy in runs 1, 2, and 3. In run 4
($\gamma=1.01$), the total energy outflow flux is dominated by the kinetic and
internal energy -- the contributions from both components are comparable. 
This suggests that the centrifugal forces may be responsible for an outflow in
run 4, rather than magnetic forces (runs 1-3). We are elaborating this point below.

\subsection{Formation of a stationary outflow} 

In this section, we investigate how the large scale bipolar
outflow is formed. We will do it in the context of previous work, namely PB03b,
\citet{proga:2003} and \citet{proga:2005}.  We will
concentrate on identifying the flow components and checking if they are by
nature the same ones as those found in previous simulations.  Therefore we
first briefly summarize the results from PB03b.

As mentioned in the introduction, PB03b considered models similar to ours but
only for $\gamma=5/3$. They found that the nature of a weakly magnetized,
slightly rotating accretion flow is controlled by magnetic fields when
$l\simgreat  2 R_{\rm S} c$.  As in the HD inviscid case, studied in PB03a (see also
MP08), the magnetized material with $l\simgreat 2 R_{\rm S} c$ forms an equatorial
torus. However, in the MHD case, the torus accretes on to BH because of 
magnetocentrifugal instability (MRI; \citealt{balbus:1991}) and
magnetic braking. The magnetic field in the torus amplifies and produces both
a magnetic corona and an outflow. The latter two can be strong enough to prevent
accretion of the low-{\it l} material through the polar regions, which was the
source of accretion in the HD inviscid case.

Very similar time evolution and flow behaviour was found in simulations of the
MHD collapsar model for gamma-ray bursts \citep{proga:2003} which were similar
to those of PB03b's simulations, except for much more sophisticated 
micro-physics. In particular, Proga et al.'s considered an equation of state
that included contributions from gas, photons and electrons. The most relevant
aspect of Proga et al.'s simulations for us here is that the inner torus is
radiation pressure dominated and the effective adiabatic index of the inner
torus is $\approx 4/3$.  A general conclusion from PB03b and 
\citealt{proga:2003} is that
for $\gamma\simgreat 4/3$ the torus formation and evolution, 
including the development of a corona
and outflow are qualitatively similar. In particular, the torus
outflow is strong enough to suppress polar funnel accretion and
to propagate out to a large distance from the BH.
This result is perhaps unsurprising
given that the torus is geometrical thick for $\gamma\simgreat4/3$.

The previous simulations showed that slightly rotating accretion flows are
very variable and complex. In particular, the inner radius of the torus and
the strength of the outflow change with time. For these reasons, only a later,
much more detailed analysis of PB03b and \citet{proga:2003} simulations showed
an additional flow component: a very collimated outflow driven from an
infalling gas with $l\simless 2 R_{\rm S} c$ \citep{proga:2005}.  To articulate the
basic physics that occurs in production of this collimated outflow,
\citet{proga:2005} performed also some new simulations. He considered gas with
$l< 2 R_{\rm S} c$ so that after an initial transient, the flow in the HD case
accretes directly on to a BH without forming a torus.  However, in the MHD
case, even with a very weak initial magnetic field, the flow settles into a
configuration with four components: (1) an equatorial inflow, (2) a bipolar
outflow, (3) polar funnel outflow, and (4) polar funnel inflow. The second
flow component of the MHD flow represents a simple yet robust example of a
well-organized inflow/outflow solution to the problem of MHD jet formation.

In each of our present simulation, the accretion flow forms a torus and
outflow although some low-{\it l} ($l< 2 R_{\rm S} c$) inflow may occur for
most of the time.  We confirm that an outflow may form in two manners.  First
type of an outflow is driven by centrifugal and magnetic forces as it was
already described in \citet{proga:2005}. The second type of outflow is a wind
driven from a torus by the toroidal magnetic field gradient as described in
PB03b (2003b and references therein).  In the first case, the outflow is
collimated and removes only a small fraction of the low-{\it l} matter from
the polar funnel  and does not affect the torus evolution, i.e. the torus
is separated from the collimated outflow by an accreting stream of low-{\it l}
matter. In the second case, 
the outflow is a broad wind from the torus.
This wind can remove all low-{\it l} matter from the polar funnel 
(by pushing it sideways). The two
types of outflows contribute the time variability of $\MDOT_{\rm a}/\MDOT_{\rm
B}$ described in the previous section and of the flow structure detailed here.

\begin{figure*}
\begin{picture}(0,600)

\put(0,450){\includegraphics{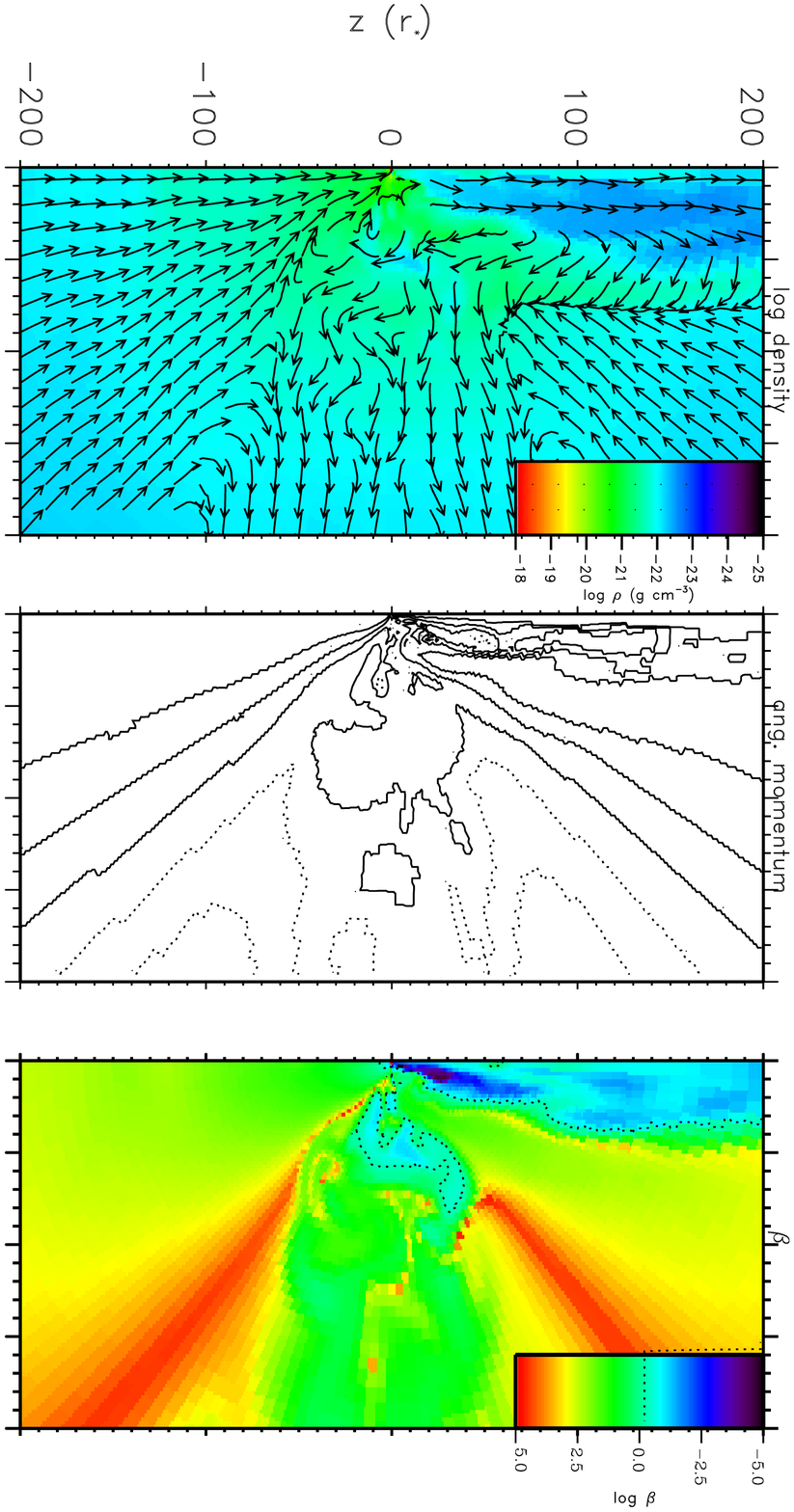}}
\put(0,300){\includegraphics{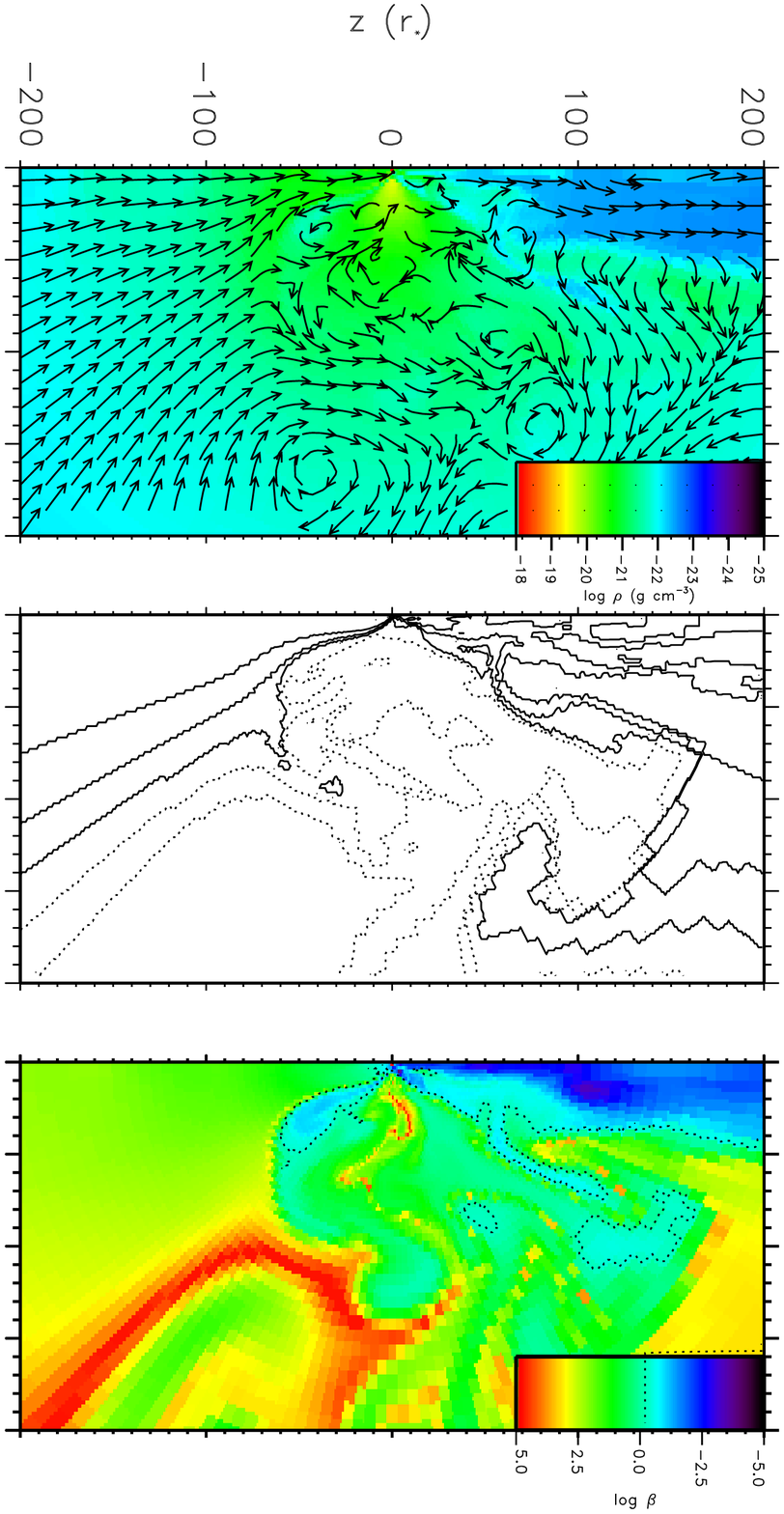}}
\put(0,150){\includegraphics{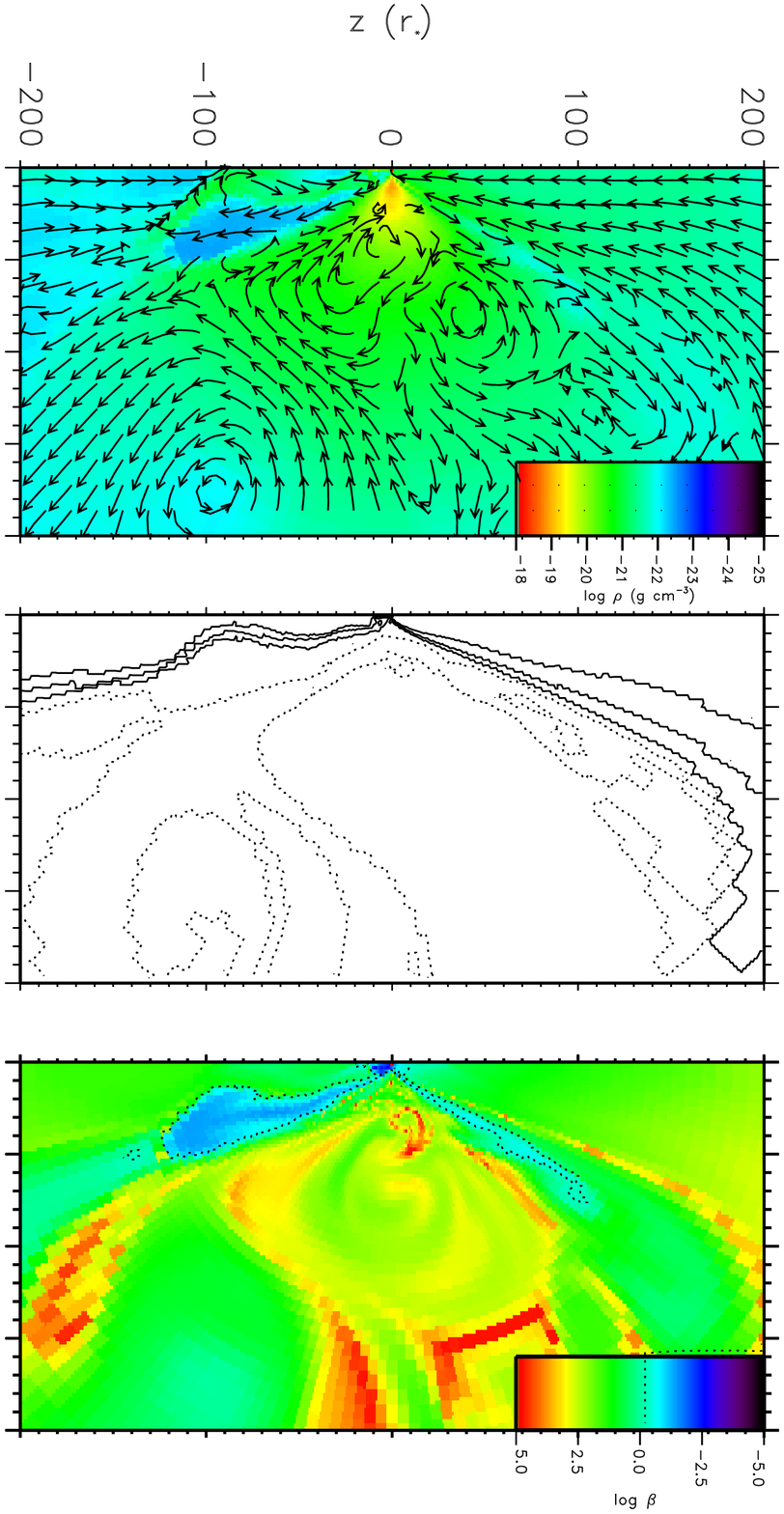}}
\put(0,0){\includegraphics{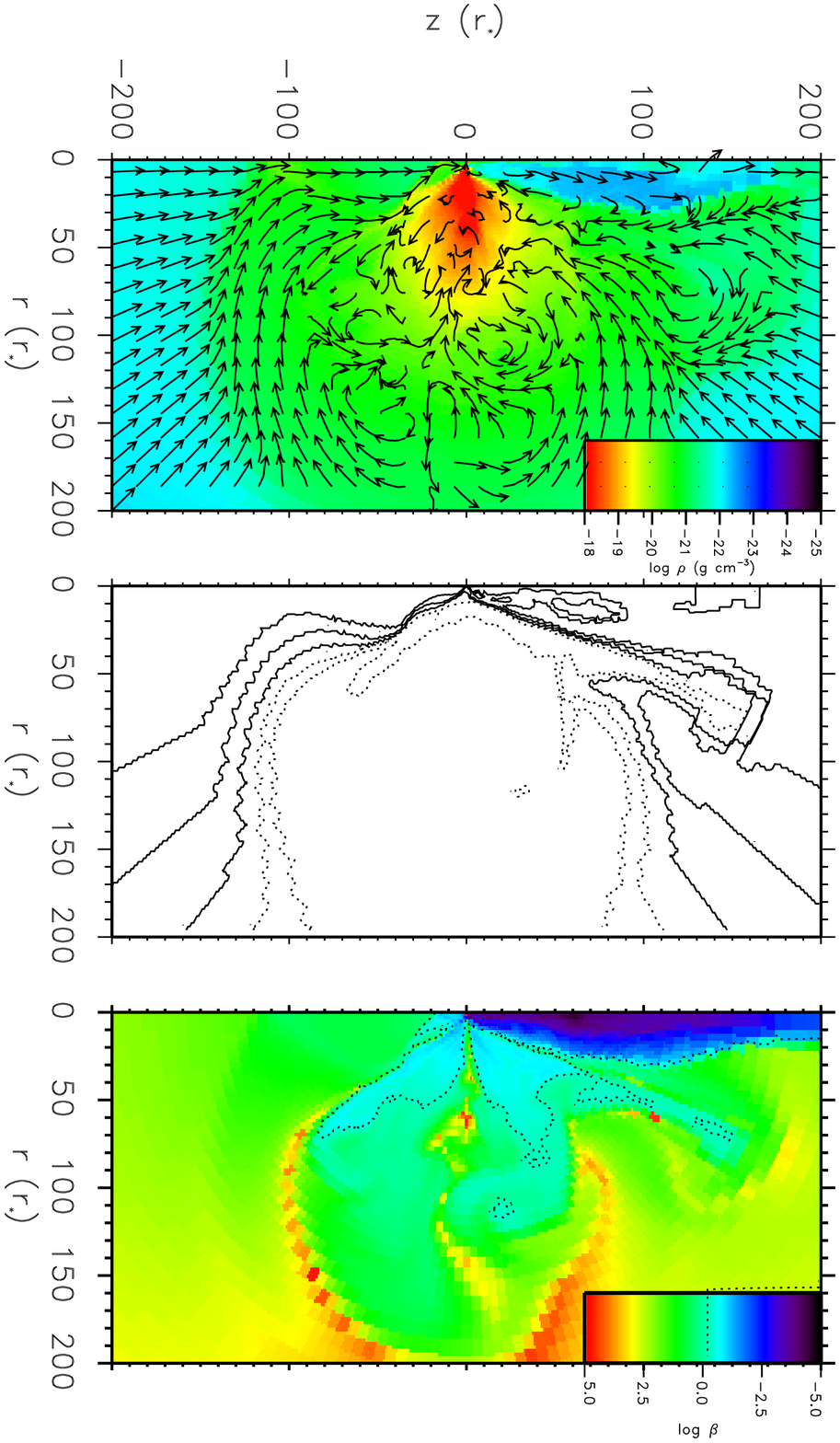}}
\end{picture}

\caption{The figure shows two-dimensional structures of the density and
velocity fields (left panels) along with the angular momentum contours
(middle panels), and the structure of
the magnetic field parameter log($\beta$) (right panels), 
at the moments right after the
`one-sided polar outflow' is formed. Each panel shows inner 200 $R_{\rm S}$.
  Panels from top to bottom show outflow
for different values of $\gamma$ for t=2800, 3400, 3450, 7500, for
$\gamma$=5/3, 4/3, 1.2, and 1.01 from top to bottom, respectively.
The angular momentum contours are range from 0.3 to 1.5 (in units of
$l_{\rm cr}$) from symmetry axis toward equator with step of 0.3.  
Solid contours indicate $l<l_{\rm cr}$, and dotted contours $l>l_{\rm cr}$.
Dotted contour of log$\beta$ shows where log $\beta$ changes from
positive to negative value.}
\label{fig:mhd_corona}
\end{figure*}

\begin{figure*}
\begin{picture}(0,600)

\put(0,450){\includegraphics{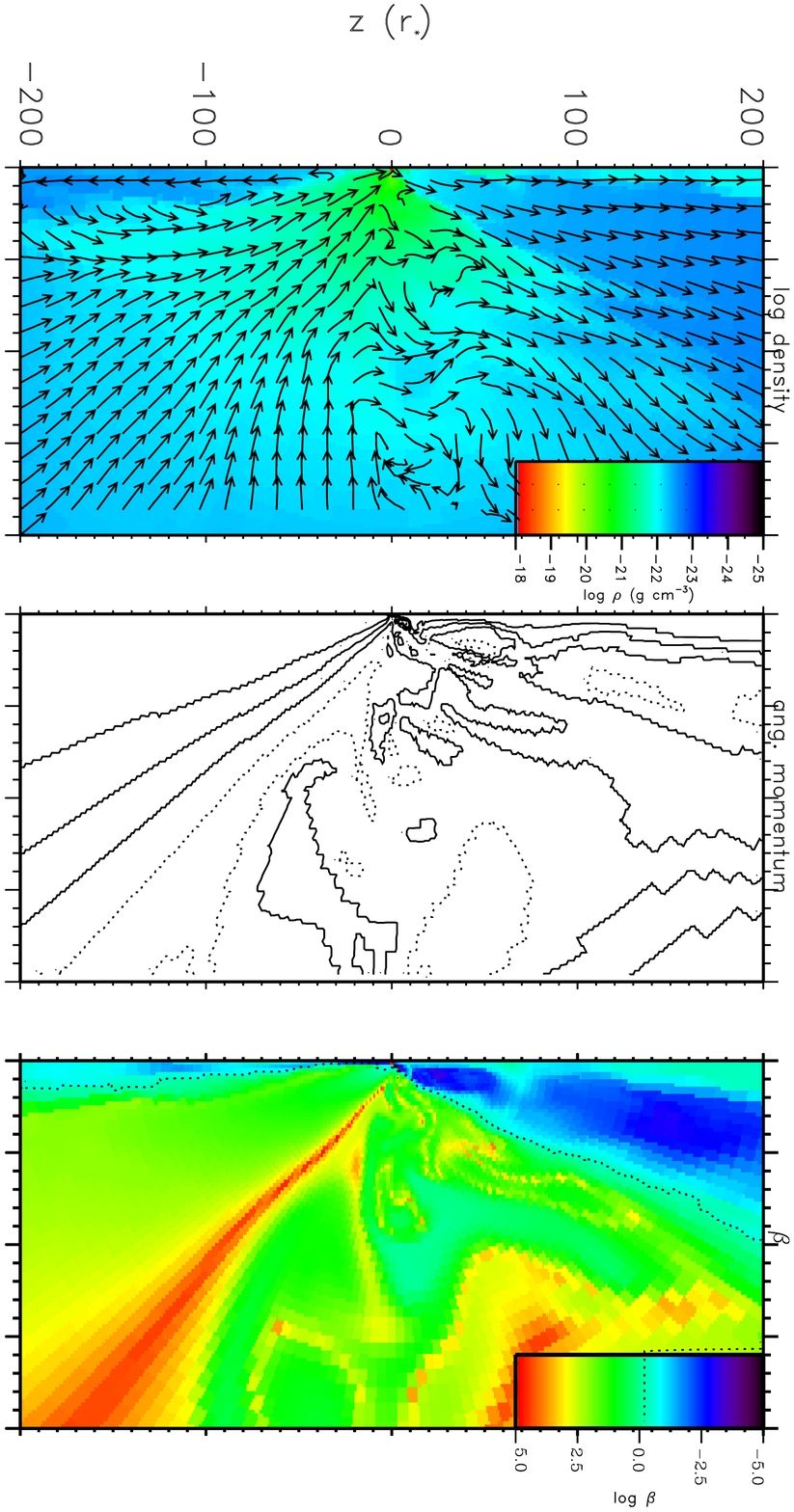}}
\put(0,300){\includegraphics{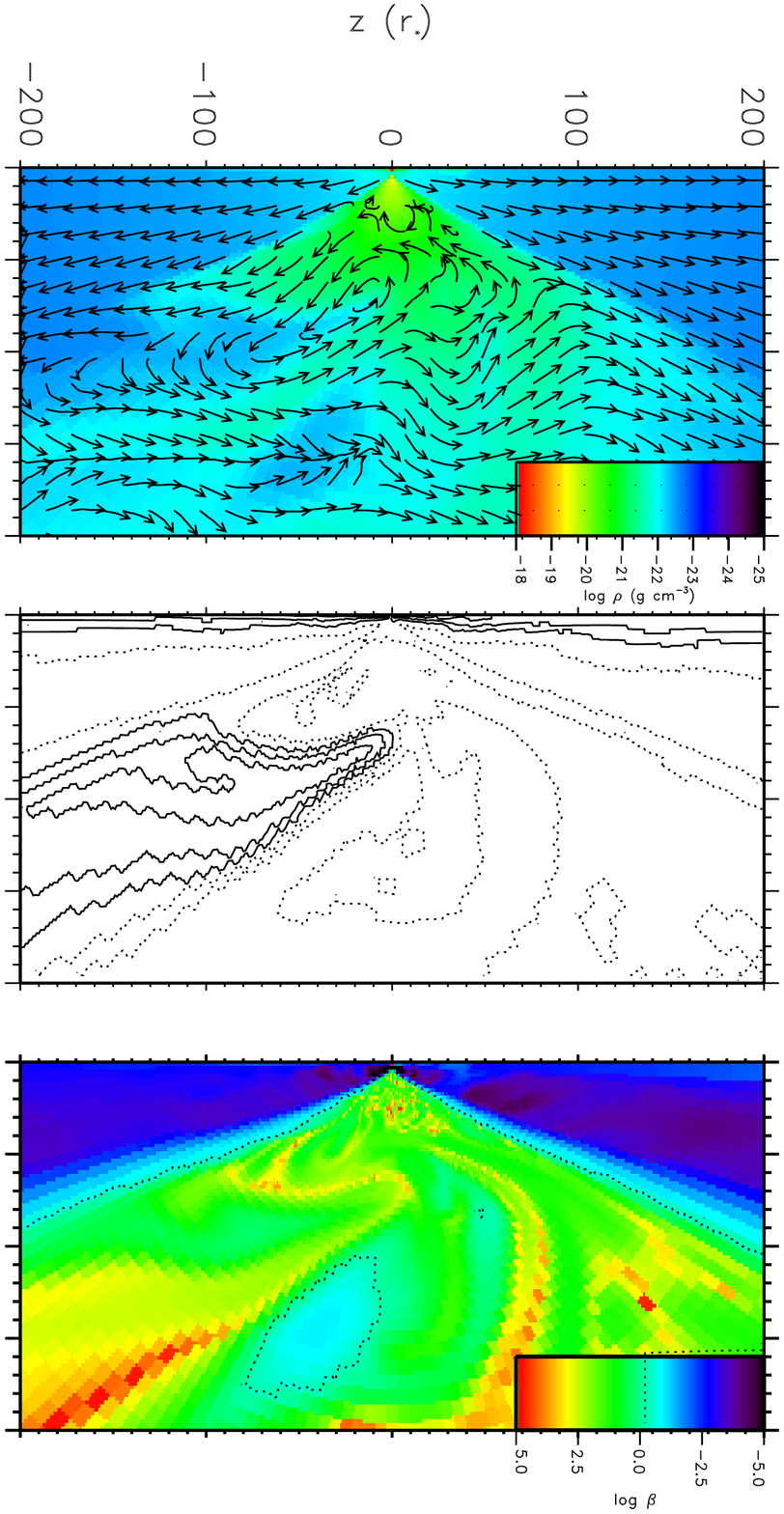}}
\put(0,150){\includegraphics{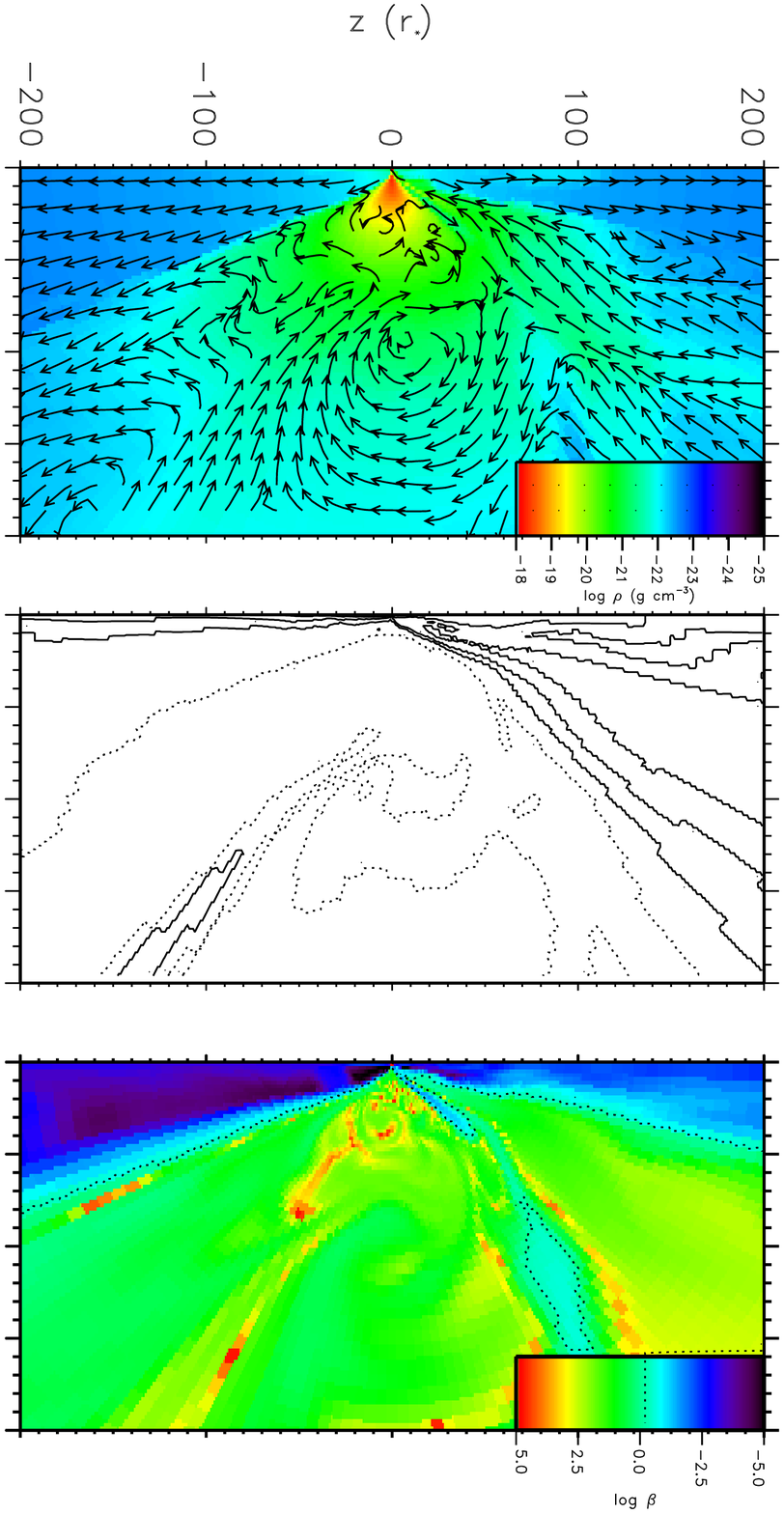}}
\put(0,0){\includegraphics{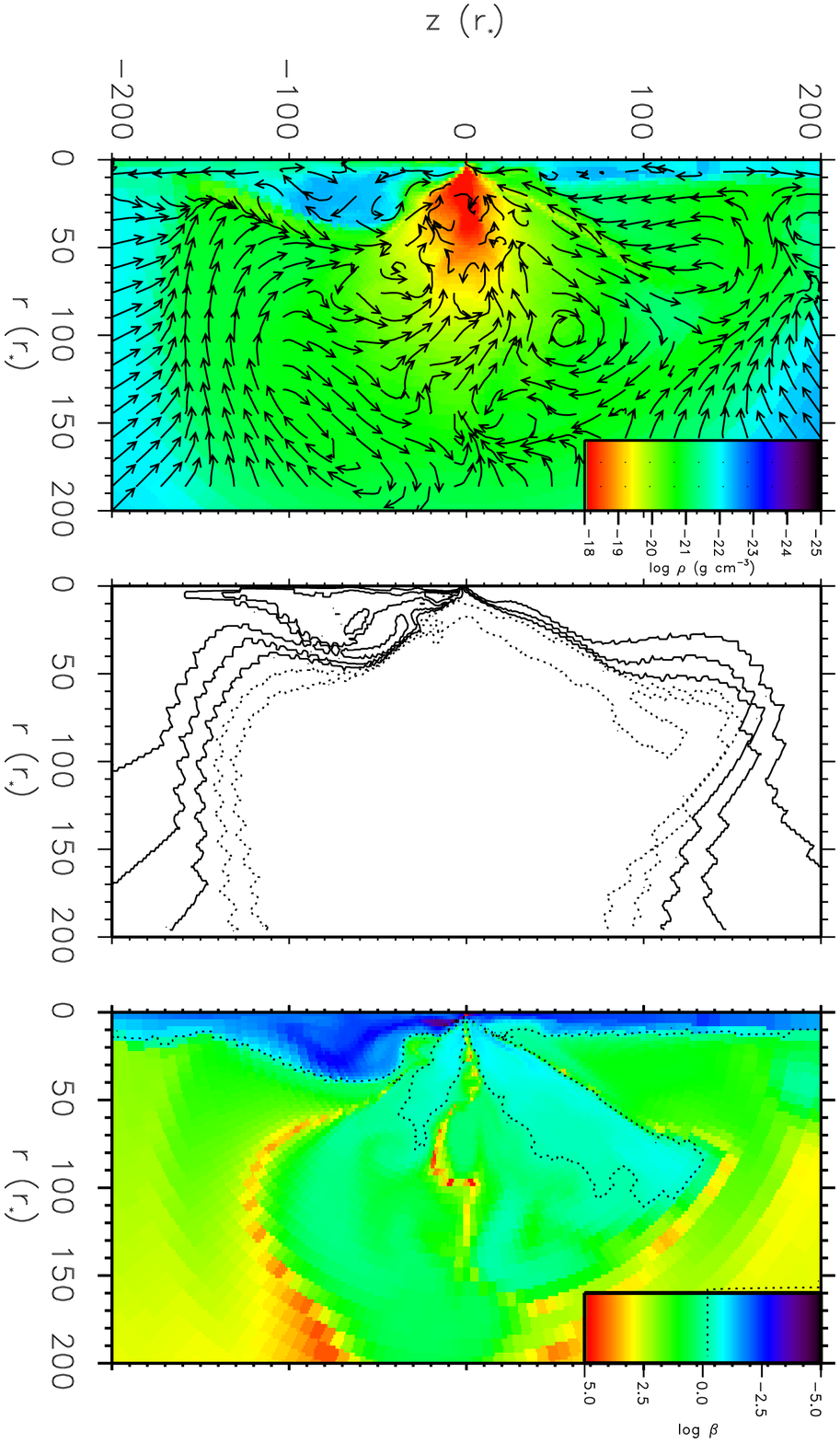}}

\end{picture}
\caption{Same as Fig.~\ref{fig:mhd_corona} at the moments right after the
polar outflow is formed on the opposite side of the torus.  Panels from top to
bottom show outflow for different values of $\gamma$ for time moments t=3500,
4000, 6000, 8300, for $\gamma$=5/3, 4/3, 1.2, and 1.01 from top to bottom,
respectively. The value of angular momentum and log($\beta$)  
as in Fig.~\ref{fig:mhd_corona}.}
\label{fig:mhd_corona2}
\end{figure*}

In Figs.~\ref{fig:mhd_corona} and~\ref{fig:mhd_corona2}, we
show the two-dimensional structures of the density and velocity fields
along with the contours of the angular momentum of the gas. We also show
the maps of the magnetic field parameter $\beta=8 \pi P/ B^2$, including total
magnetic filed ($B^2=B_r^2+B_{\theta}^2+B_{\phi}^2$), at the times of the
formation of an outflow on N and S side.

For $\gamma$=5/3 (run 1), the high-{\it l} matter ($l > 2 R_{\rm S} c$, dotted
lines on the angular momentum contours) does not reach the BH and a torus does
not form yet ( here, we define the torus as a high-{\it l}
matter accreting in the equatorial region). Instead in the inner flow, the low-$l$ gas and the
high-{\it l} gas are mixed. Such a state of accretion was not found in the
previous simulations because here we assume a continuous distribution of 
{\it l} with $\theta$ at $r_{\rm o}$, whereas in the previous simulations of PB03b
a step-function was assumed.

In run 1, the initial outflow is driven by the centrifugal and magnetic forces
along the rotation axis as found by \citet{proga:2005} (but not in PB03b).  As
we mentioned in the previous subsection $\MDOT_{\rm a}/\MDOT_{\rm B}$ does not
decrease significantly during the formation of the initial outflow
(Fig.~\ref{fig:accretion_evolution} left-upper panel), but it does decrease
slightly some time after the outflow is formed. 

The polar outflow is highly collimated by the infalling ambient gas. The
outflow that forms on the opposite side of the torus later during the
simulation is very similar to that just described. At the moment of the
outflow formation, the torus is not developed yet. The
outflow is marked by the solid contours in the angular
momentum panel in Fig.~\ref{fig:mhd_corona} (upper middle panel). As one
can see this outflow is
made of the slowly rotating matter near the symmetry axis.

In the later phases of the simulations (Fig.~\ref{fig:mhd_corona2}), after the
bipolar outflow expansion, the high-{\it l} gas forms a well developed torus
that extends inward to the last stable orbit and accretes on the BH.
Subsequently, smaller scale (within 20 $R_{\rm S}$) magnetic torus corona and
outflow develop. In this phase, the torus outflow starts to push aside the
low-{\it l} matter accreting from the polar regions. For the time up to
t$\approx$ 6000, the high- and the low-{\it l} gases mix with each other, near
the BH.  Later, when the low-{\it l} matter is significantly pushed aside by
the fully developed magnetic torus corona and outflow, the accretion flow
becomes very similar to that in PB03b (see also Sect.~\ref{sec:structure}).

For $\gamma$=4/3 (run 2, second row of panels in Figs.~\ref{fig:mhd_corona}
and~\ref{fig:mhd_corona2}), the outflow is formed in the similar way as found
by PB03b  i.e. it is a collimated wind from a torus magnetic
corona.  In this case, the torus is more bound than in the case of
$\gamma$=5/3. Therefore the high-{\it l} matter does not mix with the
low-{\it l} matter near the equator close to the BH (the upper panels of
fig. 8 in MP08
 for the HD run 1).  The magnetic torus corona and outflow
form relatively
 fast. The corona in this case extends up to $r\sim 70 R_{\rm
S}$.  Because
 the torus is geometrically thick, and the low-{\it l} matter
is pushed aside
 very efficiently from the polar region, the polar inflow is
completely
 suppressed ( see Fig.~\ref{fig:mhd_corona2} middle panel
corresponding to $\gamma=4/3$).  The moment of the sudden decrease of
$\MDOT_{\rm a}/\MDOT_{\rm B}$ (Fig.~\ref{fig:accretion_evolution}) corresponds
to the time of the outflow formation.

Next, we describe what happens for $\gamma$=1.01 ( run 4, bottom panels in
Figs.~\ref{fig:mhd_corona} and~\ref{fig:mhd_corona2}).  Already in the HD
simulations (runs G, H, I, J in fig.8 in MP08), one may notice that because of
much lower gas pressure, the torus is geometrically thinner compared to the
cases with higher $\gamma$. Here, a strongly turbulent magnetic corona forms
above the disc. The corona is geometrically thick, but it does not expand into
the polar regions as in the previous cases. As in run 1, initially the outflow
is caused by the centrifugal and magnetic forces on both sides (as in
\citealt{proga:2005}).  However, this collimated outflow does not completely
prevent the low-{\it l} matter inflow toward the BH from the polar
regions. Therefore, $\MDOT_{\rm a}/\MDOT_{\rm B}$ remains always high (close
the Bondi rate). In the lower panels in Figs.~\ref{fig:mhd_corona}
and~\ref{fig:mhd_corona2}, the dotted contours in the angular momentum panels
indicate that the high-{\it l} matter is separated from the outflow by the
low-{\it l} accreting matter marked as the solid contours. The low-$\beta$
corona is separated from the low-$\beta$ collimated outflow with the
high-$\beta$ region of the accreting matter.  As we mentioned in the previous
section, the moment of the outflow formation cannot be clearly identified in
the mass accretion rate curve (Fig.~\ref{fig:accretion_evolution}). The
persistently accreting low-{\it l} matter separates the polar outflow and
magnetic corona for most of the simulation time.  In this case, the accretion
flow consists of four parts (from the equator to the pole): (1) geometrically
thin torus in the equatorial plane, (2) highly turbulent magnetic corona, (3)
low-{\it l} matter inflow through polar funnels, and (4) polar outflow.

In the intermediate case of $\gamma$=1.2 (run 3, third row of panels in
Figs.~\ref{fig:mhd_corona} and~\ref{fig:mhd_corona2}), although the magnetic
corona forms, the polar outflow forms in both manners as described for runs 1
and 2. In Fig.~\ref{fig:mhd_corona}, one notices the corona formation on the N
and S sides.  The corona extends beyond $r = 100 R_{\rm S}$, which is clearly
seen in Fig.~\ref{fig:mhd_corona}.  First, on the S pole, the expanding corona
causes the outflow. The outflow on the opposite pole, is caused by the
magneto-centrifugal mechanism described in \citet{proga:2005}. The mass
accretion rate, just as in the case of $\gamma$=1.01, does not experience a
sudden reduction during its evolution because the outflow is weak so that the
low-{\it l} matter persistently accretes and mainly contributes to the total
mass accretion rate. Nevertheless, the mass accretion rate for $\gamma$=1.2 is
slightly lower than that for $\gamma$=1.01.

In our models, the collimation of the jet-like outflow depends on the
$\gamma$ index and it is related to the mechanism of its formation rather than
the torus geometrical thickness. For $\gamma=5/3, 4/3$ and $\gamma=1.2$ (S
side), the jet-like outflow at the final time of the simulation is not well
collimated, because it is driven from the torus corona. 
To investigate the problem of collimation, one should perform
the simulation over a longer time period to allow the flow to reach 
the quasistationary state at larger radii. 
For $\gamma=1.01$ and $\gamma=1.2$ (N side) the outflow is more collimated in
comparison to the coronal jet-like outflow because
it is formed by the low-{\it l} matter and collimated by the accreting
low-{\it l} matter.

\subsection{Episodic coronal outflows for low $\gamma$}
\label{sec:outburst}
For $\gamma$=1.01 (as well as for $\gamma$=1.2), $\MDOT_{\rm a}/\MDOT_{\rm B}$
remain high for most of the simulations time.  However, at the late phases of
the simulations, we observe a few significant minima in the mass accretion
rate curves (lower panels in Fig.\ref{fig:accretion_evolution}).  These minima
are caused by small scale coronal outflows which temporary suppresses
accretion of the low-{\it l} matter in the polar funel.

\begin{figure*}
\begin{picture}(0,600)

\put(100,330){\includegraphics{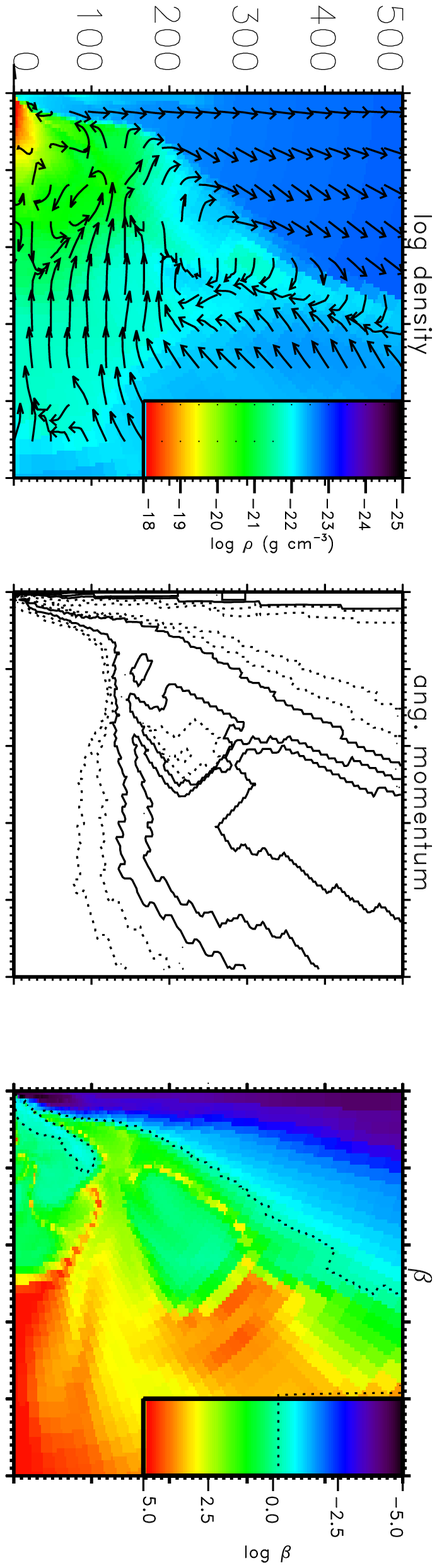}}

\put(100,220){\includegraphics{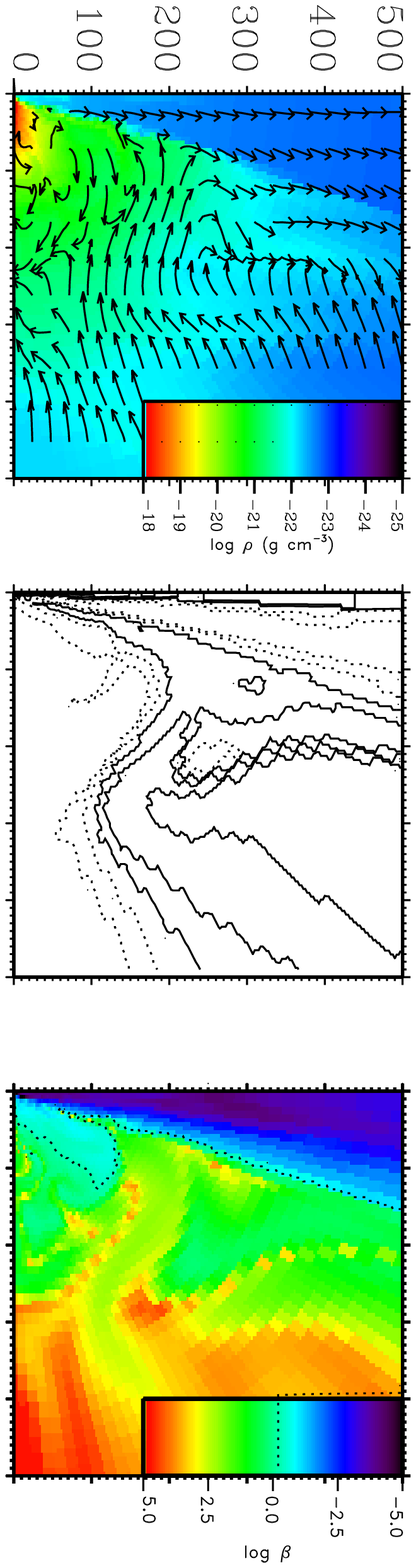}}

\put(100,110){\includegraphics{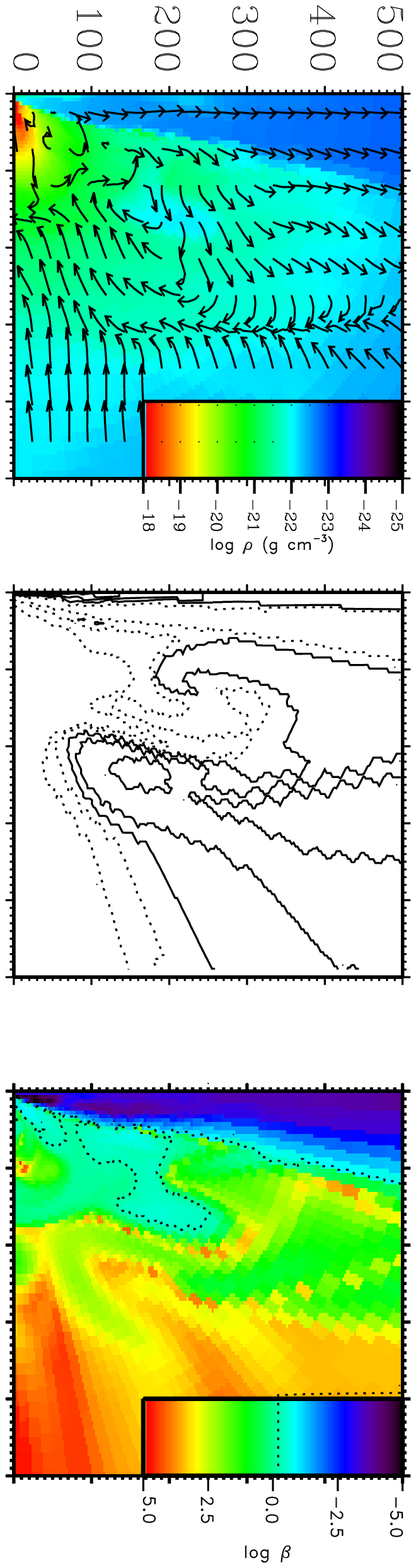}}

\put(100,0){\includegraphics{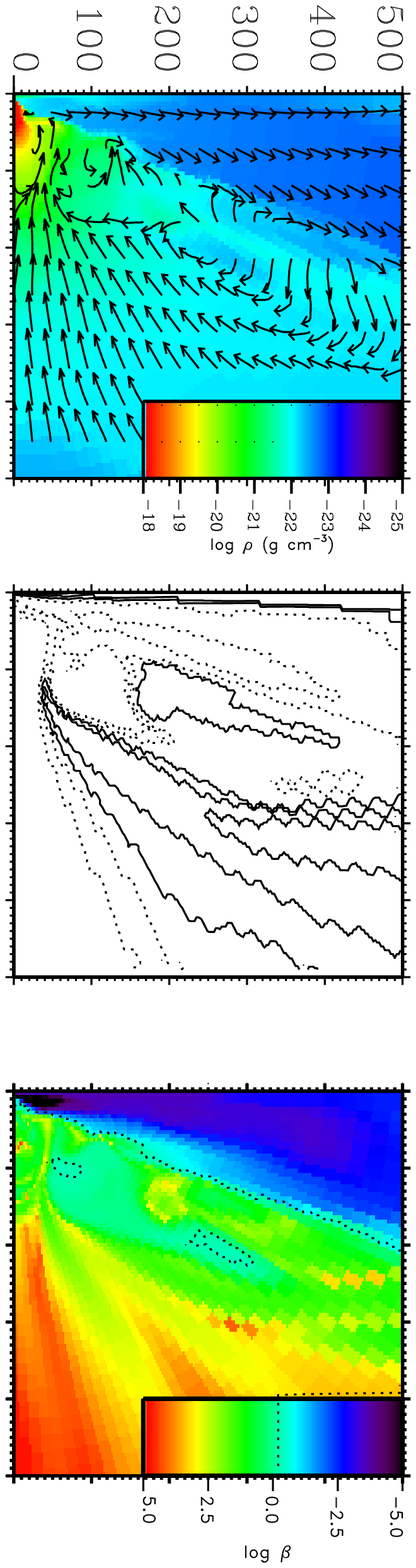}}

\put(100,-110){\includegraphics{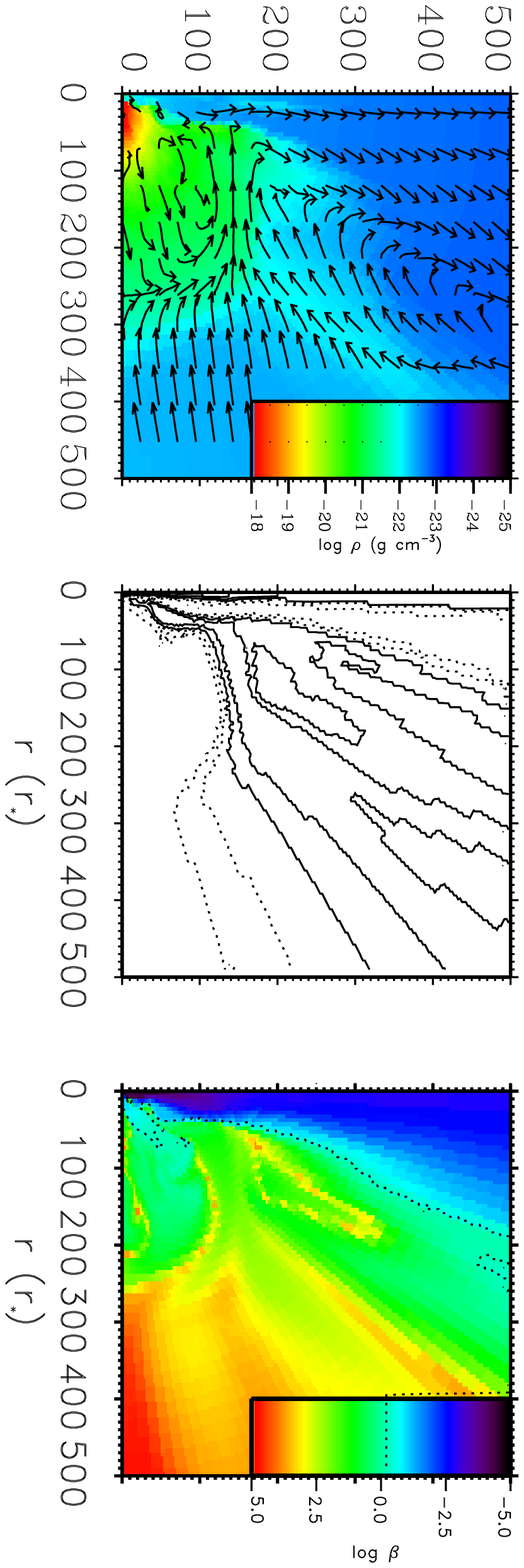}}
\end{picture}
\caption{The evolution of density overploted with velocity field, 
specific angular momentum contours, and plasma parameter $\beta$
are shown during an coronal outburst in run 4. The time increases
from upper to bottom panels, t=9500,10000,10500,11000,13000 $t_{\rm dyn}$.}
\label{fig:burst}
\end{figure*}

Fig.~\ref{fig:burst} presents the sequence of maps of the logarithmic density,
angular momentum contours, and plasma parameter $\beta$, before, during, and
after an episode of an outflow in run 4.  To better show the structure of the
inner flow, we present only a half of the computational domain above the
equator within 500 $R_{\rm S}$.  Before the outflow, the magnetic torus corona
extents to 100 $R_{\rm S}$ (the two top rows of panels).  During an outflow,
the corona expands up to 300 $R_{\rm S}$ (the third and fourth rows of
panels).  Later, the matter flows back toward the BH.  Before and after the
outflow, the torus corona and polar outflow are separated by a stream of
the low-{\it l} matter. After the outburst, $\MDOT_{\rm a}/\MDOT_{\rm B}$
increases back to its mean high level (the bottom row of panels).  The whole
cycle lasts about 3500 $t_{\rm dyn}$.

\subsection{The structure of the torii}
\label{sec:structure}

\begin{figure*}
\begin{picture}(0,600)
\put(-100,300){\includegraphics{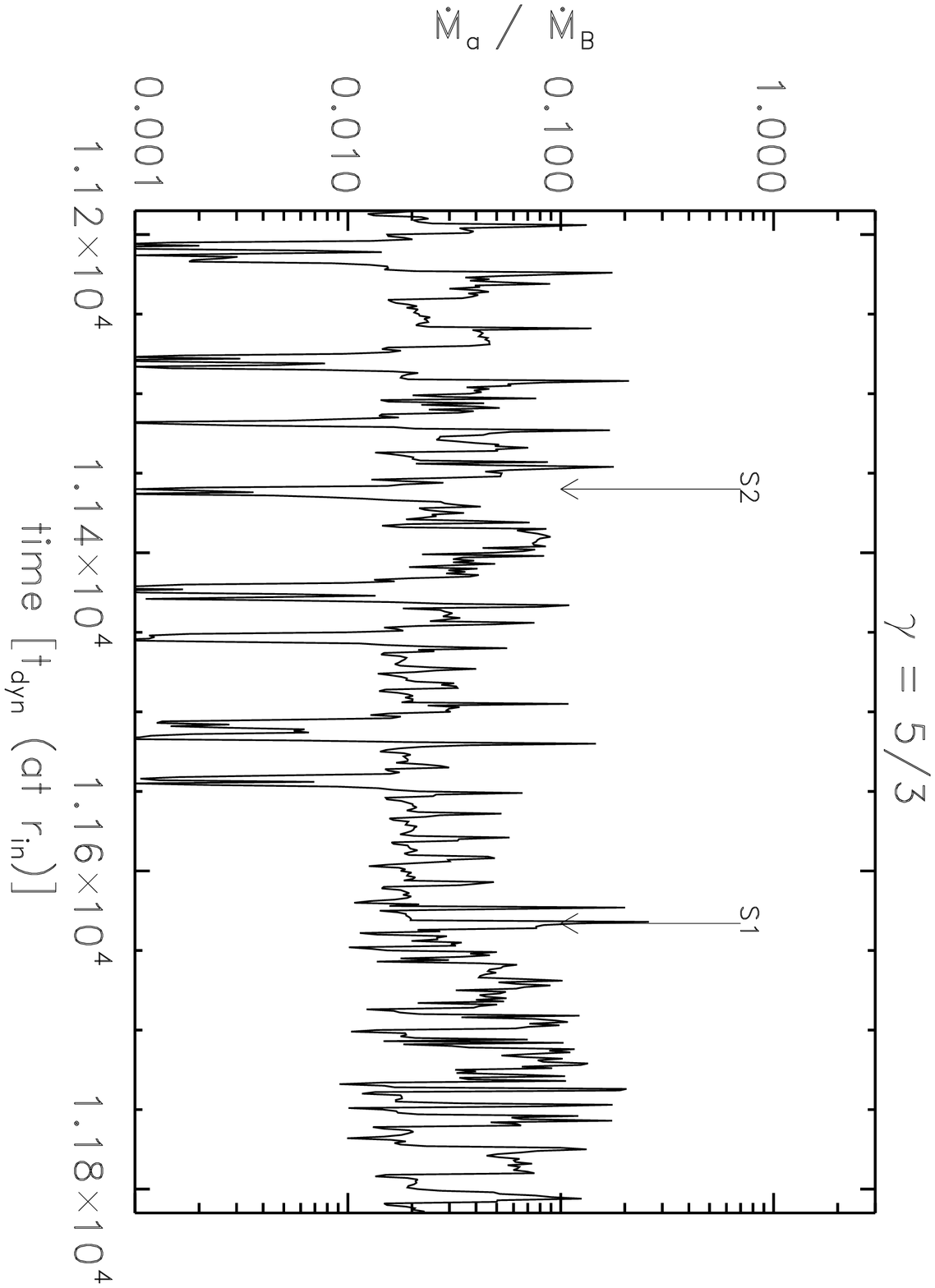}}
\put(140,300){\includegraphics{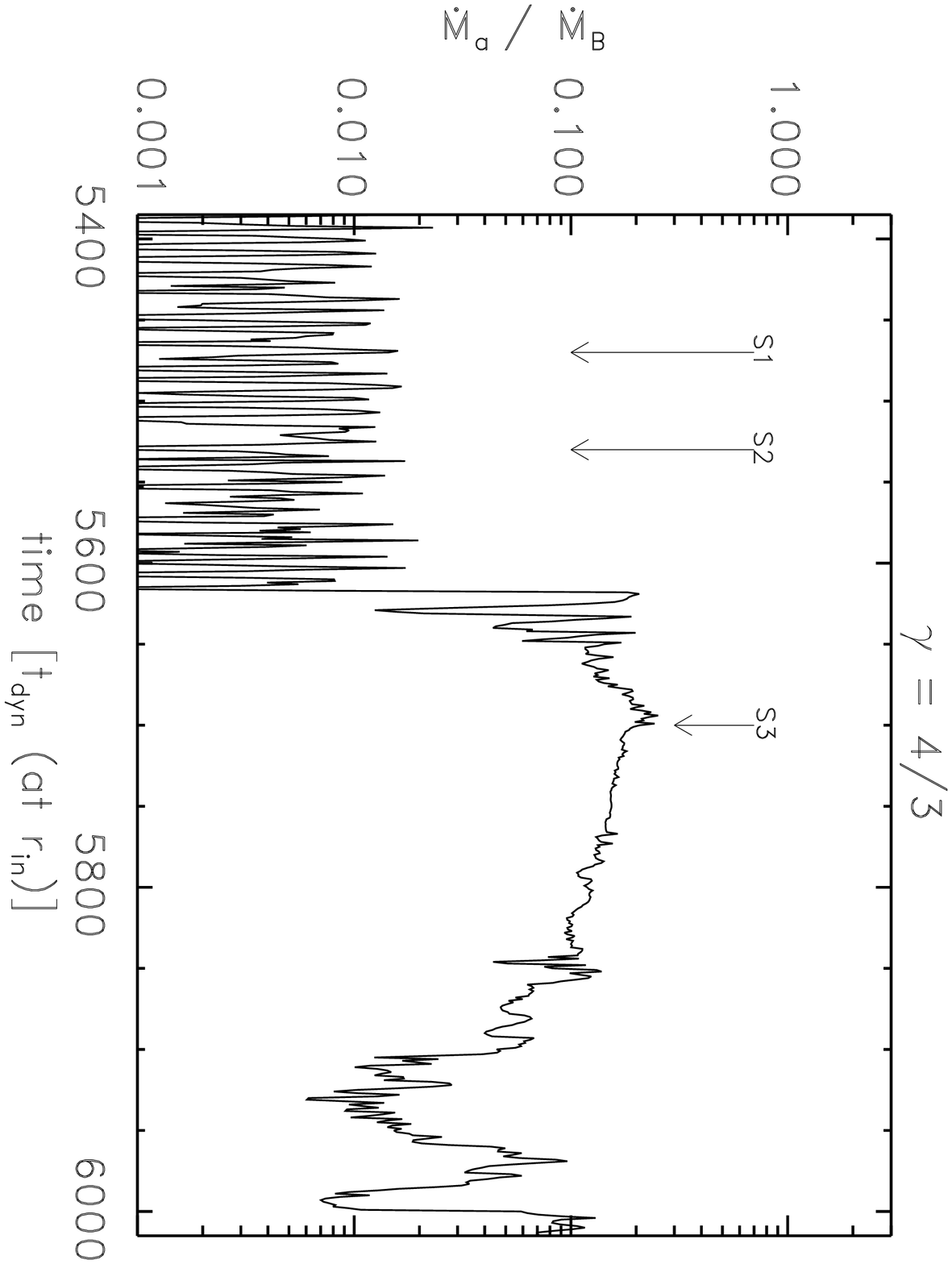}}
\put(-100,100){\includegraphics{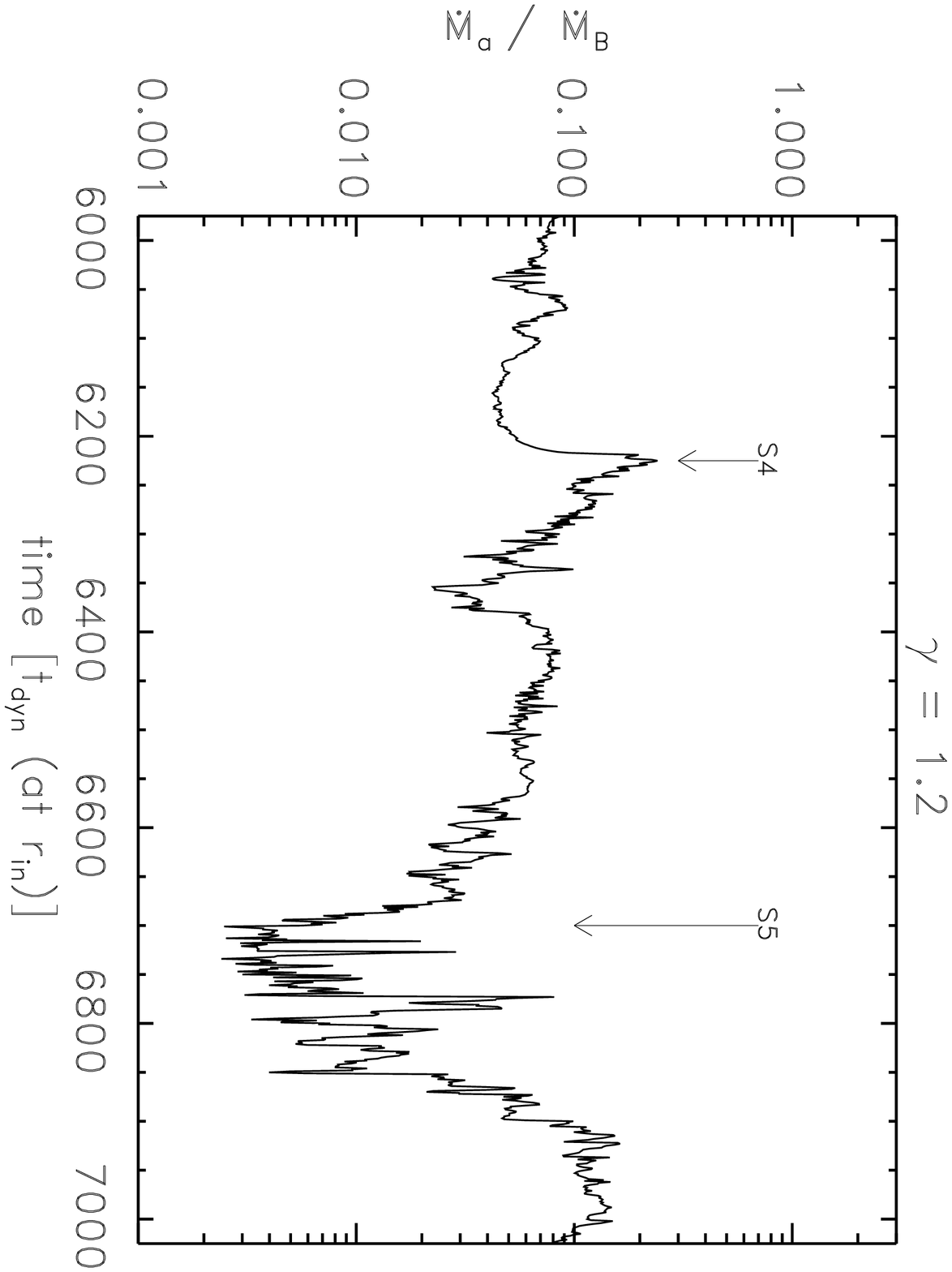}}
\put(140,100){\includegraphics{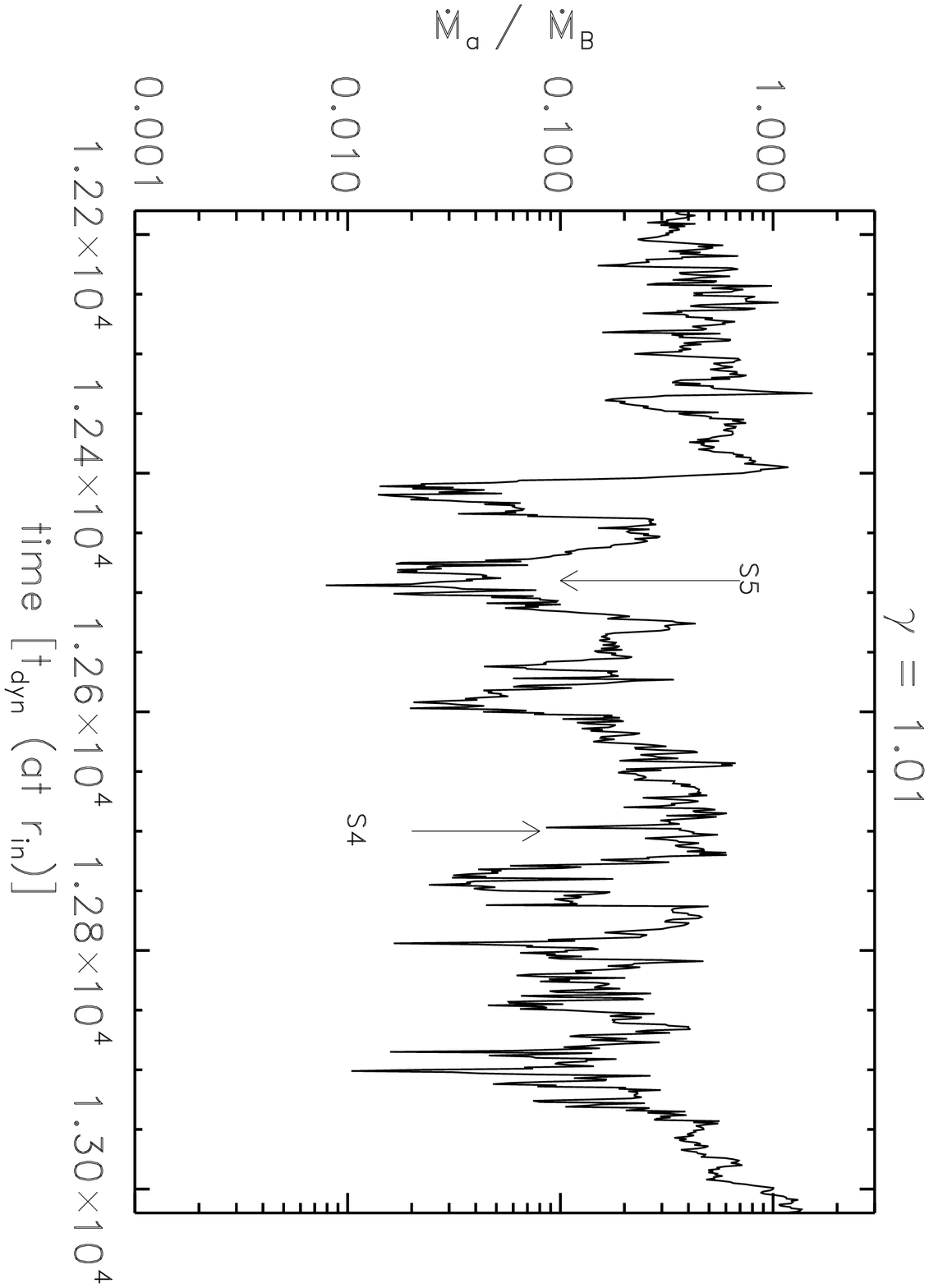}}
\end{picture}
\caption{Four panels show the evolution of the accretion rate 
for different $\gamma$ parameter. $t_{\rm dyn}$=595 s.}
\label{fig:accretion_evolution_mini}
\end{figure*}

In this section, we return to our investigation of the time evolution of
$\dot{M}_{\rm a} / \dot{M}_{\rm B}$ for various $\gamma$. Here, we identify
accretion states that correspond to various episodes of torus accretion.
Figs.~\ref{fig:accretion_evolution_mini} shows the segments of the $\MDOT_{\rm
a}/\MDOT_{\rm B}$ evolution for runs 1, 2, 3, and 4, which contain the typical
`accretion states' that appeared for a given model. The characteristic times,
are marked with arrows denoted as S1, S2 (in run 1 and 2), S3 (only in run 2), 
S4 and S5 (in run 3 and 4).

We consider the flow to be unstable (MRI) when the following relation
holds \citep{balbus:1998}:
\begin{equation} 
u_A^2 < - \frac{h^2}{\pi^2} \frac{d \Omega^2}{d ln r},
\label{eq:disk_thickness}
\end{equation} 
where $u_A$ is the Alfv{\'e}n speed, $\Omega$ is the angular velocity, 
$r$ is the distance from the
BH and {\it h} is the half of the disk thickness. 
We define {\it h} as the height at which the
density of the matter decreases by {\it e} along the vertical direction in
comparison to the density at the equator. 
For run 1, 2, and 3, the disk height can be approximated
with its radius h=r in Eq.~\ref{eq:disk_thickness}. 
For run 4, we adopt the numerical value of $h$ ($h$ roughly corresponds to 1/3 r).
In the vertical direction, 
the torus typically contains $\sim50, 42, 40$ and 24 grid points, for 
$\gamma$=5/3, 4/3, 1.2, and 1.01, respectively. 
In the radial direction, there are $\sim 60$ and $\sim 90$
equaly spaced in logarithmic scale 
zones within first 20 $R_S$ and 100 $R_S$, respectively.
To find, whether the fastest growing mode
of the MRI is resolved with our
computational mesh, we compute its critical wavelength 
$\lambda_{crit}$ as defined by eq.~113 in \citet{balbus:1998}. 
We checked the ratio between $\lambda_{crit}$ and $\Delta x$ (the
grid spacing) as a function of position on the grid.  If the ratio is larger
than one, the grid resolves torus for MRI. In regions where the ratio is
smaller than one the grid does not resolve the critical wavelength
$\lambda_{crit}$.

We have analysed the flow properties at each accretion state as PB03b did.
For $\gamma$=5/3, we found that the accretion state S1 corresponding to the
narrow peak in Fig.~\ref{fig:accretion_evolution_mini} (panel for
$\gamma$=5/3) appears when the high-{\it l} torus accretes on to the BH.
We recognize S1 as `B' state found in PB03b. During S1 the
torus accretes because of the magneto-rotational instability (MRI). We checked that
the current grid does resolve the fastest mode of the MRI.
$\lambda_{crit} / \Delta x > 10 $ within $r < 40 R_S$.
The accretion at the peak is dominated by the accretion from torus, 
($\dot{M}_{\rm a} /\MDOT_{\rm B}$ reaches 0.06) and typically stays at this level for 
less than $\sim 5.1$ $t_{\rm dyn}$. S1 appears in the mass accretion rate
curve many times (i.e. all spikes correspond to the torus accretion).

The accretion from torus is subsequently accompanied by the polar funnel inflow
(`leaking') of the matter that was left in the polar regions after the outflow.
The density in the polar funnel is
relatively low in comparison to the torus density (or in comparison to the
low-{\it l} matter removed from the polar funnel by the outflow).  Its value is
artificially set by the assumed $\rho_{\rm min}$ parameter (see
Sect.~\ref{sec:init_mhd}). The matter from the polar funnel contributes to the
mass accretion rate and sets its mean level around 0.02
(Fig.~\ref{fig:accretion_evolution_mini}).

During state S2 in run 1, $\MDOT_{\rm a}/\MDOT_{\rm B}$ is very low and it corresponds
to a state when the torus is truncated from the BH. The truncation is due
to magnetic effects (ie., a `magnetic barrier') as in `C' state described in
PB03b (see also \citealt{proga:2006} and fig. 1 there).  
 We find that the poloidal magnetic field is larger
than the toroidal component for $r < 10 R_{\rm S}$ during this state. The magnetic
field lines are stretched for $r < 10 R_{\rm S}$.  The MRI does not operate in the
inner $20 R_{\rm S}$.  The matter from the polar funnel also flows outward during
state S2. Therefore, $\dot{M}_{\rm a} / \dot{M}_{\rm B}$ drops much below
the mean value of $\approx$ 0.02. The duration of S2 state is comparable to the
duration of S1 and all moments with $\dot{M}_{\rm a} / \dot{M}_{\rm B} < 0.02$
correspond to the same S2 state.

In our run 1, we do not see an accretion state corresponding to the `D' state
found in PB03b. In the `D' state, $\MDOT_{\rm a}/\MDOT_{\rm B}$ should
increase above 0.1 level, because of the slowly rotating matter previously
pushed away by the coronal outflow, reentering the BH vicinity.  We attribute
lack of the 'D' state in run 1 to a relatively short duration of our
simulations.  We performed our simulation for much shorter time
(Table~\ref{tab:mhd_sum}) in comparison to PB03b (2.5 $t_{\rm dyn}$).  In our
simulations, the low-{\it l} matter is far away ($r > 500 R_{\rm S}$) from the
center at the end of the simulation.

For $\gamma$=4/3, S1 state corresponds
to the episode of torus accretion, but has a much lower amplitude than narrow
peaks found for $\gamma$=5/3
( S1 is representative of all narrow peaks with amplitude of
$\dot{M}_{\rm a} / \dot{M}_{\rm B} \sim 0.01$ in
Fig.~\ref{fig:accretion_evolution_mini}, panel for $\gamma$=4/3).  
We find that MRI causes the torus accretion,
but our grid resolution was not adequate to capture the fastest growing mode
of MRI,  $\lambda_{crit} / \Delta x < 1.5 $ within $r < 40 R_S$.
 Thus, the mass accretion rate from the torus is lower than it would
be if the fastest mode of MRI were resolved.  In S1 state, the torus accretion
is accompanied by accretion of the matter with density of $\rho_{\rm min}$
from the polar region.  This polar funnel accretion sets the mean accretion
rate at the level of 0.01, similar to the mean accretion rate obtained for
$\gamma$=5/3.

During S2 state (Fig.~\ref{fig:accretion_evolution_mini}, panel for
$\gamma$=4/3), the torus is pushed away from the BH by the magnetic
barrier (as in S2 state for $\gamma$=5/3 case).

The mentioned `D' state appears in run 2 despite of a short duration of the
simulation. S3 state (Fig.~\ref{fig:accretion_evolution_mini}, panel for
$\gamma$=4/3) is characterized by the the sudden increase of $\MDOT_{\rm
a}/\MDOT_{\rm B}$ up to the level of $\sim 0.2$ and much longer duration $\sim
300 t_{\rm dyn}$ compared to state marked as S1. In
Fig.~\ref{fig:accretion_evolution_mini}, panel for $\gamma$=4/3,
one can see only one state S3, 
but in Fig.\ref{fig:accretion_evolution} one can find two broad
features corresponding to state S3. 
During state S3, the low-{\it l} matter accretes on
to the BH near the equator.  This matter has been pushed away from the BH by
the expanding magnetic corona during the bipolar outflow formation but not as
far as in run 1.  Therefore the weakly rotating matter may reenter the region
close to the BH in a much shorter time-scale.

In the intermediate case of $\gamma$=1.2 (run~3), the mass accretion rate
remains relatively high during the whole simulation 
(Fig.~\ref{fig:accretion_evolution_mini}, panel for $\gamma$=1.2) due to
almost persistent low-{\it l} polar radial inflow. 
Therefore, contrary to run 1 and 2, it is difficult to distinguish 
any characteristic patterns in the mass accretion rate curve.
S4 marks $\MDOT_{\rm a}/\MDOT_{\rm B}$ peaking at 0.2, when
the torus accretes along with the low-{\it l} matter. 
However, in this case the grid does not
resolve the fastest growing mode of the MRI for $r < 20 R_{\rm S}$.  
The torus accretion is dominated by the slower MRI modes. The main contribution 
to $\dot{M}_{\rm a}$ comes from a polar inflow of the low-{\it l} matter.
The mass accretion rate is slightly higher than average because the
matter from polar regions simultaneously accretes from both, N and S sides
(S4).  State S4 appears only a few times during run~3. For most of the time,
the matter in the polar regions accretes from one side only.
S4 is similar to the S3 in run~2, but during S3 low-{\it l} matter approaches
the BH in the equatorial plane rather than from the poles, like in S4.

In S5 state, the mass accretion rate is dominated mainly by the torus, because
almost all matter is removed from the polar funnel due to the episodic coronal
outburst described in \S~\ref{sec:outburst}.  At that time $\dot{M}_{\rm
a}/\dot{M}_{\rm B}$ becomes lower than 0.01.  In run~3, the high-{\it l} torus
never undergoes the truncation because the poloidal magnetic field is very
weak and does not form a barrier for accretion.  For $\gamma=1.2$ S5 appears
two times at the end of the simulation. Note that S5 is similar to state
S1 in run~1 and 2, but here the torus accretes inefficiently, therefore S5 
appears as a dip  (not as a spike like in e.g. run~1) 
above the mean level of accretion.

For $\gamma$=1.01 (run~4), $\dot{M}_{\rm a}/\dot{M}_{\rm B}$ remains
relatively high (Fig.~\ref{fig:accretion_evolution_mini}, panel for
$\gamma$=1.01) for most of the duration of the simulation. The low-{\it l}
matter accretes quite persistently (as in run~3) along the torus surface (S4),
on its N or S sides, or both.  Although MRI operates in the torus, again the
accretion fastest mode is not resolved with the simulation. Thus, the torus
accretion rate is underestimated. In run 4, the torus never experiences a
truncation. The low-{\it l} matter inflow is only occasionally stopped by the
turbulent corona but such episodes are short since the outflow is weak (S5).

\section{Discussion and Conclusions}
\label{sec:4}

In this article we present results from the axisymmetric
2-D MHD simulations of a low angular
momentum accretion flow on to the BH. Our work is an extension of MP08,
which in turn was initiated by PB03a and PB03b 
(see also \citealt{proga:2003}). Here, we concentrate on 
investigation of effects of gas pressure on the flow structure
and evolution. In particular,
we study how the flow
properties change as a function of the adiabatic index, $\gamma$.  
We also compare our MHD runs to their HD counterparts presented in MP08. 

We find that in the early phase of the evolution, the MHD runs behave very 
similar to those in MP08. Namely, the matter with $l > l_{\rm cr}$ forms a torus 
near the equator. As expected,  the torus thickness decreasing with decreasing
$\gamma$. The similarity in the MHD and HD initial evolution, is a consequence 
of our assumption of a initially very weak magnetic field. At later times,
however, the MHD torii do not persistently accumulate 
matter near the BH as their HD counterparts do but rather accrete on the BH 
due to MRI and magnetic braking. 
In addition, the MHD torii are far more turbulent and 
also slightly geometrically thinner than their HD counterparts.
However, most importantly contrary to the HD torii, 
the MHD torii produce bipolar outflows.

In MP08, we noticed that for $\gamma=1.01$,
the torus is excited by the inflowing material and shows small amplitude 
quasi-periodic oscillations in the direction perpendicular
to the equator.  In the MHD runs, 
we do not observe clear torus oscillations in the power spectrum of the mass 
accretion rate or in any other way . This difference is due to the fact
that in the MHD simulations the
high-{\it l} torus is more turbulent and does not have a well
defined outer boundary what would separate it clearly from the inflow
(such a boundary is very clear in the HD simulations -
see the bottom row of panels fig. 9 in MP08).

One of the properties of the innermost torus that 
is common for the HD and MHD simulations
is the cylindrical distribution of the angular velocity in
torus for all $\gamma$.

The bipolar outflow and angular momentum transport by magnetic
effects transforms the structure of the flow from a
polar-inflow/equatorial outflow (or equatorial accumulation) in the HD models 
into a polar-outflow/equatorial accretion in the MHD models.  
However, the torus outflow 
is strong only in models with $\gamma$=5/3, 4/3 and 1.2.
For $\gamma$=1.01, the outflow is very weak and cannot propagate 
throughout the inflow.

In the MHD runs with a strong torus outflow, the total mass accretion rate of 
the gas on to the BH is reduced because $\MDOT_{\rm a}$ is dominated by the polar
funnel accretion of the low-{\it l} gas which the torus outflow suppresses.
The mass accretion rate is also more variable in the MHD simulations  than
in the HD ones. This variability is caused mainly by mixing of
the low- and high-{\it l} gas. As we showed in MP08, the degree of the
mixing increases with the distances from the center
for $\gamma$=4/3 and 1.2.
The mixing causes episodes of an enhanced polar funnel accretion
in the HD models (bursts). In the MHD models, 
we observe similar bursts not only for $\gamma$=4/3 and 1.2 but also for 
$\gamma$=5/3.  In both HD and MHD models, $\MDOT_{\rm a}$ can increase by a factor 
of few.
For $\gamma$=1.2 in the MHD run, we do not observe very
strong flares seen in the HD counterpart.

Our new simulations for $\gamma$=5/3 
are consistent with the previous results obtained by PB03b.
We found however, some minor differences between PB03b and 
our simulations. These differences occur during the initial phase of the
evolution 
which we attribute to the differences in the initial conditions
described in \S~\ref{sec:2}. In addition,
we do not observe here the episodes of enhanced accretion of the low-{\it l} 
matter (call by PB03b the accretion state D). This difference is due to 
a shorter duration of our simulations compared to PB03b's simulations  
(see the previous section).

The torus outflow is not the only outflow we find in our simulations, i.e.,
we see also a highly collimated outflow driven by the centrifugal and magnetic 
forces. As found  by \citep{proga:2005}, this
outflow appears regardless of the high-{\it l} torus formation 
and  is produced from the low-{\it l} inflow
(see \citealt{proga:2003}). The `outflow-from-inflow' does not 
significantly influence 
the torus evolution.  If the matter in the polar region were not rotating 
(e.g. a $\theta$ distribution of {\it l}
were a step function at the outer boundary) 
this outflow would not form as it was the case in PB03b's simulations.

The most important new result in the present simulations is that for a
geometrically thin disc model ($\gamma \approx 1$) the torus outflow is
strongly suppressed. Generally, the structure of the torus and its magnetic 
corona  depend on $\gamma$.
For $\gamma$=5/3, gas pressure is significant and the flow is hot and
the torus corona extends far from the equator.  
For $\gamma$=1.01, the magnetic corona above the disc is
highly turbulent, and it is small and gravitationally bound. The
magnetic field in the corona is less tangled in the flows for $\gamma
\simless 5/3$ than the flow with $\gamma$=1.01.

As mentioned above, the torus outflow changes the mass accretion rate.
For $\gamma$=5/3, $\MDOT_{\rm a}/\MDOT_{\rm B}$
dropped suddenly about one order of magnitude
after the outflow develops. 
This drop in the accretion rate is due to depleting of 
the polar funnel. We observe the same behaviour of $\MDOT_{\rm a}/\MDOT_{\rm B}$ 
in the run with $\gamma$=4.3. For $\gamma$=1.2, the outflow
is significantly weaker and is often one-sided. 
For $\gamma$=1.01, the torus outflow is very weak and is stalled by the inflow.
The mass accretion
rate remains high for the most of the simulation time in comparison to PB03b
runs and our runs for $\gamma$=5/3 and 4/3.

In runs 1 and 2, we observe the accretion state 'C' 
which occurs when the torus is truncated  by the magnetic
barrier. This state was first identified by PB03b. However,
in the runs 3 and 4 ($\gamma$=1.2 and 1.01) we observe the stationary torus 
accretion (called by PB03b as state B) with the state C never occurring. 
This indicates that for low $\gamma$, the role of the magnetic field is 
reduced  not only in producing a torus outflow but also in changing the torus
accretion. 
We note that the reduced role of magnetic fileds may be
also due to resolution of our models. The MRI cannot be resolved
with $140 \times 100$ points on mesh for $\gamma < 5/3$, because the disk
thickness decreases with $\gamma$ (half of the disk thickness $h \sim 2/3 r$ for
$\gamma=4/3$ and 1.2 and $h \sim 1/3 r$ for $\gamma=1.01$). We plan perform
simulations with higher resolution in the near future.

The direct comparison of the simulations with the observations of BHB
and AGN spectra are beyond the scope of this work. Therefore, 
we can only make some remarks on the observational consequences of our models.  
For example, the strong torus outflow for $\gamma \simless 5/3$ and the
suppression of the outflow $\gamma$=1.01 may be relevant to the observed
transition from the hard/low state to the soft/high state. 
The episodic outburst found in models with $\gamma$=1.2 and 1.01, may
correspond with the flares seen in a very high states in BHB.
To relate our results with observations, 
we plan to carry out calculations of broad-band spectra predicted by
our models following \citet{moscibrodzka:2007}.

Our models are axisymmetric, which may affect the results. In axisymmetry we
cannot model the toroidal field MRI and a poloidal
magnetic field can  dissipate according to the antidynamo theorem
\citep{moffatt:1978}. We note that even in the HD limit, the axisymmetric models
differ from non-axisymmetric ones. For example, \citet{janiuk:2008} found 
that in fully 3-D simulations the inner torus processes. 
It would be important to 
check if the torus will also precess in the MHD limit and what consequences 
it would have on the outflow formation and the polar-funnel accretion.  
Recent work on outflows driven by radiation from a precessing
disc show that the flow geometry can be dramatically changes
although the mass accretion rate can be similar to that
in the non-precessing disc case \citep{kurosawa:2008}. 
We plan to address these issues in our future work.

Future simulations should also include radiative cooling and heating processes.
Recently, \citet{fragile:2009} pointed out that the radiative cooling may 
be of the same order as artificial cooling caused by lack of the
treatment of kinetic and magnetic dissipative
processes in non-conservative codes (as one used by us). 
Our models also do not resolve the geometrically thin discs.  Currently, 
global simulations of the magnetized geometrically thin disc 
is beyond computational abilities.

We acknowledge support provided by the Chandra awards TM7-8008X (M.M. and
D.P.) and TM8-9004X (D.P.) issued by 
the Chandra X-Ray Observatory Center, which is operated by 
the Smithsonian Astrophysical Observatory for and on behalf of 
NASA under contract NAS 8-39073.


\newpage

\clearpage

\newpage

\clearpage

\newpage

\newpage

\newpage

\newpage

\clearpage
\newpage

\end{document}